\def\tPm{\tilde{\bm P}_m}
\def\tPs{\tilde{\bm P}_s}

\def\Pm{{\bm P}_m}
\def\Ps{{\bm P}_s}
\def\Pns{{\bm P}_{\underline s}}
\def\eps{\varepsilon}
\def\half{\tfrac{1}{2}}
\def\trho{\tilde{\rho}}
\def\us{{\underline{s}}}

\newcommand{\vectornorm}[1]{\left|\left|#1\right|\right|}
\documentclass [twocolumn,superscriptaddress]{revtex4}
\usepackage {graphicx}
\usepackage {amsmath}
\usepackage {bm}
\usepackage {subfigure}

\newsavebox{\mylegend}
\sbox{\mylegend}
{\centering
\includegraphics[width=0.12\textwidth,clip]{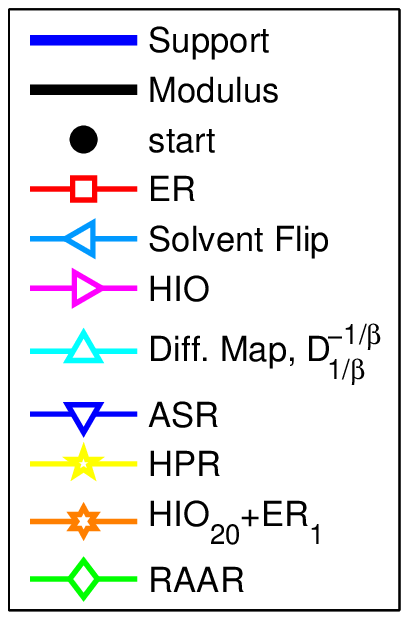}
}

\begin{document}

\title{A unified evaluation of iterative projection algorithms 
for phase retrieval}

\begin{abstract}
Iterative projection algorithms are successfully being used as a
substitute of lenses to recombine, numerically rather than optically,
light scattered by illuminated objects. Images obtained
computationally allow aberration-free diffraction-limited imaging and
the possibility of  using radiation for which no lenses exist.
The challenge of this imaging technique is transferred from the lenses
to the algorithms.  We evaluate these new computational
``instruments'' developed for the phase-retrieval problem, and discuss
 acceleration strategies.
\end{abstract}

\author{S.~Marchesini}
\email[Correspondence and requests for materials should be addressed
to S. Marchesini: ]{smarchesini@lbl.gov}
 \affiliation{Lawrence Livermore National Laboratory, 7000 East Ave.,
Livermore, CA 94550-9234, USA}
 \affiliation{
 Center for Biophotonics Science and Technology, University of California, Davis, 2700 Stockton Blvd., Ste 1400, Sacramento, CA 95817, USA}
\altaffiliation[Current address: ]{Lawrence Berkeley National Laboratory, 1 Cyclotron Rd, Berkeley CA 94720, USA.}



\date{\today}
\maketitle

Crystallographers routinely image molecular structures of several
thousand atoms by phasing the diffraction pattern of a structure
replicated in a periodic system. Likewise, computationally retrieving
the phase of a diffraction pattern is becoming increasingly successful
at imaging -- with several millions of resolution elements-- 
objects as complex as  biological cells,
nanotubes and nanoscale aerogel structures. Diffraction microscopy
(the imaging of isolated objects by diffraction and computational
phase retrieval) promises a 3D resolution limited only by radiation
damage, wavelength, the collected solid angle and the number of x-rays
or electrons collected.  This capability provides an extremely
valuable tool for understanding nanoscience and cellular biology.
Recent estimates
\cite{Howells:2004} of the dose and flux requirements of x-ray
diffraction on single objects indicate that attractive resolution
values (about 10 nm for life science and 2--4 nm for material science)
should be possible at a modern synchrotron. Atomic resolution could be
accomplished using pulses of x-rays that are shorter than
the damage process itself \cite{solem, Neutze:2000} using femtosecond pulses from
an x-ray free-electron laser \cite{chapman:2006np}.  Alternatively 
the radiation damage limit could be eliminated by
continuously replacing the exposed samples,  such as
laser-aligned molecules \cite{Spence:2004} with identical ones.

In the fields of electron microscopy \cite{Gerchberg:1972} and
astronomical imaging \cite{Fienup:1978}, iterative projection
algorithms have been used to recover the phase information in a
variety of problems.  The evaluation of the aberrations in the Hubble
space telescope described by Fienup in \cite{Fienup:hubble} 
remains perhaps the
most prominent example of successful phase reconstructions in the
astronomical community. Nugent and collaborators applied similar techniques to
characterize x-ray lenses \cite{Nugent:2006}. In electron diffraction microscopy
\cite{Gerchberg:1972, spence:2002}, Zuo and coworkers imaged a single
isolated nanotube at atomic resolution
\cite{zuo}, Wu et al. imaged defects at atomic resolution 
\cite{Wu:nmat}.

An important review, which attempted to integrate the approaches of
the optical and crystallographic communities, appeared in
1990 \cite{Millane:1990}. The connection was made between the
``solvent-flattening" or ``density-modification" techniques of
crystallography \cite{itc} and the compact support requirements of the 
iterative projection algorithms. 
 The importance of fine sampling of the
\textit{intensity} of the measured diffraction pattern was recognized 
at an early stage \cite{Bates:1982}.

The observation by Sayre in 1952 \cite{Sayre:1952} that Bragg
diffraction undersamples the diffracted intensity pattern was
important and led to more specific proposals by the same author for
x-ray diffractive microscopy of non-periodic objects
\cite{Sayre:1980, Miao:1998}. These ideas, combined with the rapid
development of computational phase retrieval in the wider optics
community, especially the ``support constraint''
\cite{Gerchberg:1972,Fienup:1978,Fienup:1980,Fienup:1982}, 
enabled the first successful use of coherent x-ray diffraction microscopy (CXDM).  

Since the first proof of principle demonstration of CXDM by a team at Stony Brook
\cite{Miao:1999}, a number of groups have been working to bring these
possibilities into reality.

Robinson and co-workers at the University of Illinois have applied the
principles of CXDM to hard x-ray experiments on microcrystalline
particles. Such data have been reconstructed tomographically to
produce a 3D image at 80 and more recently 40 nm resolution \cite{Williams:2003, Pfeifer:2006}. Miao (now
at UCLA) and co-workers made considerable progress in pushing the
CXDM method at Spring-8 Japan to higher resolution in 2D (7 nm),
higher x-ray energies, and to a limited form of 3D
\cite{Miao:2006}. They have also made the first application of CXDM to
a biological sample \cite{Miao:2003}.

A diffraction chamber dedicated to diffraction microscopy
\cite{Beetz:2005} has been used to image biological cells
\cite{Shapiro:2005, Lima:2005} at the Advanced Light Source in
Berkeley \cite{Howells:2002}. Using the same chamber, a collaboration
between Berkeley and Livermore labs and Arizona State University produced
3D imaging at $10\times10\times40$ nm resolution of test samples
\cite{Chapman:2006} as well as aerogel foams \cite{barty}.

In this article the computational instruments that enabled these and
other results are reviewed.  Section \ref{sec:1} introduces the phase problem
and the experimental requirements for diffraction microscopy, Section
\ref{sec:2} describes the concepts of sets of images and their projectors. In
Section \ref{sec:IPA} the iterative projection algorithms published in the
literature are summarized and tested on simple geometric sets.  In
Section
\ref{sec:gradients} the connection between projection- and gradient-
based methods and related acceleration strategies are discussed.

\section{The phase problem}\label{sec:1}
When we record the diffraction pattern intensity scattered by an object, the
phase information is missing.  Apart from normalization factors, an
object of density $\rho(\bm{r})$, $\bm{r}$ being the coordinates in
the \textit {object} (or \textit {real}) space, generates a
diffraction pattern equal to the modulus square of the Fourier
transform (FT) $\tilde \rho(\bm{k})$:
\begin{eqnarray}
\nonumber
I(\bm{k})&=&|\tilde \rho(\bm{k})|^2 \\
I(\bm{k})&=&\tilde \rho^\dagger (\bm{k}) \tilde \rho (\bm{k})\,,
\label{eq:phase}
\end{eqnarray}
where $\bm k$ represent the coordinate in the Fourier (or Reciprocal)
space.  The inverse Fourier transform (IFT) of the measured intensity
$I$ provides the autocorrelation $\rho(-\bm{r})*\rho(\bm{r})$ of the
object:
\begin{equation}
 \mathrm{IFT} [ I(\bm{k})]= \rho(-\bm{r})*\rho(\bm{r})\,.
\label{eq:acor}
\end{equation}
The phase-retrieval problem consists of solving $\tilde \rho$ in
Eq. (\ref{eq:phase}) or $\rho$ in Eq. (\ref{eq:acor}), using some extra prior
knowledge. In diffraction microscopy, solving such problem is performed
with giga-element large-scale optimization algorithms, described in
the following section.

Since the intensity represents the FT of the autocorrelation function,
and the autocorrelation is twice as large as the object, the diffraction
pattern intensity
should be sampled at least twice as finely as the amplitude to capture all 
possible information on the object. Finer sampling
adds a 0-padding region around the recovered autocorrelation function,
which adds no further information (Shannon theorem).  Less than
critical sampling in the Fourier domain causes aliasing in the object
space.  A periodic repetition of the same structure provides a stronger
signal, enabling the measurement of the diffraction pattern before the
structure is damaged. However, while an isolated object generates a
continuous diffraction pattern that can be sampled as finely as
desired, a periodic repetition of the same object generates only a
subset of the possible diffraction intensities.  Crystallography
therefore has to deal with an aliased autocorrelation function, also
known as the Patterson function. This reduced information can be
compensated by other prior knowledge, such as the atomic nature of
the object being imaged, knowledge of a portion of the object,
presence of heavy atoms, and information obtained with anomalous
diffraction. Other information includes the presence of a solvent in the
crystal. By varying the sampling rate of a diffraction pattern it was
shown \cite{Millane:1996,Miao:1998,oversampling1} that less than 
critical sampling was sufficient
to solve the phase problem.  This was possible because the number of
equations (measured intensities in Eq. (\ref{eq:phase})) in the two- and
three-dimensional phase-retrieval problems is larger than the number of
unknowns (resolution elements in the object). The number of unknowns
defines the number of independent equations, or the minimum required
sampling rate. 
Although no general proof has been provided that
limited sampling removes only redundant equations, such a minimum
required sampling rate suggests that when the solvent exceeds 50\% of
the crystal volume, the algorithms developed in the optical community,
using techniques to dynamically refine the solvent regions
\cite{Marchesini:2003} may be able obtain ab-initio structural
information from crystals. 

Coherence is required to properly sample the FT of the autocorrelation
of the object \cite{Spence:coherence}. According to the Schell theorem
\cite{Goodman}, the autocorrelation of the illuminated object obtained
from the recorded intensity is multiplied by the complex degree of
coherence.  The beam needs to fully illuminate the isolated object, and
the degree of coherence must be larger than its autocorrelation.

Diffraction microscopy solves the phase problem by using the
knowledge that the object being imaged is isolated; it is assumed to
be 0 outside a region called support $S$:
\begin{equation}
\rho(\bm{r})= 0  \text {,   if  $\bm{r}\notin S$} \,.
\label{eq:support}
\end{equation}
This support is equivalent to the solvent in crystallography.
Equations (\ref{eq:phase}) and (\ref{eq:support}) can be combined to
obtain a multidimensional system of quadratic equations in the
$\rho(\bm{r})$ variables:
\begin{eqnarray}
\left |\sum_{\bm{r}\in S} \rho(\bm{r}) \exp(i \bm{k \cdot r}) \right |^2=I(\bm{k})\,,
\end{eqnarray}
which is a quadratic equation in the $\rho(\bm r)$ variables
with coefficients $c_{\bm{r,r'}}(\bm k)=\cos(\bm{k} \cdot (\bm{r-r'}))$:
\begin{eqnarray}
\sum_{\bm{r},\bm{r'}\in S}  c_{\bm{r,r'}}(\bm k) \rho(\bm{r})\rho^{*}(\bm{r'}) &=&I(\bm{k}) \,.
\label{eq:ellipsoid}
\end{eqnarray}
Each value of $I(\bm k)$ in reciprocal space defines an ellipsoid
(Eq. (\ref{eq:ellipsoid})) in the multidimensional space of the unknowns
$\rho(\bm r)$, $\{\bm r \in S\}$.  If the number of {\it independent}
equations equals the number of unknowns, the system has a single
solution $\rho(\bm r)$. The intersection of these ellipsoids forms our
solution. Unfortunately this system of equations is difficult to
solve, and has an enormous number of local minima.  Constant phase
factors, inversion with respect to the origin (enantiomorphs), and
origin shifts $\rho(\bm {\pm r+r_0})e^{i\phi_0}$ are undetermined and
considered equivalent solutions.  The presence of multiple
non-equivalent solutions in two- and higher- dimensional phase retrieval
problems is rare \cite{Barakat}; it occurs when
the density distribution of the object can be described as the
convolution of two or more non-centrosymmetric distributions. Simple
homometric structures for which the phase problem is not unique
\cite{Buerger} exist in nature, but
such non-uniqueness is less likely for more complex structures.

The presence of noise and limited prior knowledge (loose constraints) 
increases the number of solutions within the noise level and constraints.
Confidence that the recovered image is the correct and unique one 
can be obtained by repeating the phase-retrieval process using several 
random starts. Repeatability of the
recovered images as a function of resolution measures the effective
phase-retrieval transfer function \cite{Chapman:2006, Shapiro:2005}, 
which can be decomposed in unconstrained
amplitudes modes \cite{Shapiro:2005} and phase aberrations 
\cite{Marchesini:XRM_CT}.


In the early 1980s, the development of iterative algorithms with feedback 
by Fienup, produced a remarkably successful
optimization method capable of extracting phase information
\cite{Fienup:1978,Fienup:1982, cederquist:1988}. The important
theoretical insight that these iterations may be viewed as projections
in Hilbert space \cite{stark:1984,stark:1987} has allowed
theoreticians to analyse and improve on the basic Fienup algorithm
\cite{elser:2003, luke:1,luke:2,luke:3}. 

These algorithms try to find the intersection between two sets,
typically the set of all the possible objects with a given diffraction
pattern (modulus set), and the set of all the objects that are constrained
within a given area or support volume  (or outside a solvent region in
crystallography).  The search for the intersection is based on the
information obtained by \textit{projecting} the current estimate on the two
sets. An error metric is obtained by evaluating the distance between
the current estimate and a given set. The error metric and
its gradient are used in conjugate-gradient (CG) -based methods such as
SPEDEN \cite{speden}.

\section{Sets, Projectors and Metrics}\label{sec:2}
An image of a density distribution can be described as a sequence of
$n$ pixel values. For an image of $n$ pixels, there are $n$
coordinates.  The magnitude of the density at a pixel defines the
value of that coordinate. Thus a single vector in this $n$-dimensional
space defines an image. For complex images the number of coordinates
increases by a factor of 2. Axes of the multidimensional space are
formed by any sequence of $n$-pixels with all but one pixel equal to
0. An example is $ \bm{x}=(x,0,0)$ in 3-pixel solution space. The origin
of this space is the image with all the pixels equal to 0.  The
components on these axes form the {\it real} or {\it object} space.
The same object can be described in terms of any another $n$-dimensional
orthogonal (or linearly independent) bases.  Axes can be rotated,
shifted, inverted and so on, and the proper linear transform must be
applied to obtain the components in the new basis.  The basis used to
describe the image must have at least $n$ components, but more can be
used if it helps to describe the properties of the algorithm. For
example the values could be left to have a real and an imaginary
component, doubling the number of dimensions used to describe the
object.

One important basis is the {\it momentum} or {\it Fourier}
space. While the vector in the $n$-dimensional space representing an
image is unaltered on transforming from real to reciprocal space, its
components in the new axes are altered (Fourier-transformed). 
The distance between two points in the $n$-dimensional space is independent of 
this transformation (Parseval theorem).  The lengths and the
angles between vectors will be our guide to describe the behavior,
convergence and error metrics of these algorithms.

\begin{figure}
\subfigure[]
    {
      \label{subfig:support}
	\includegraphics[width=0.2\textwidth]{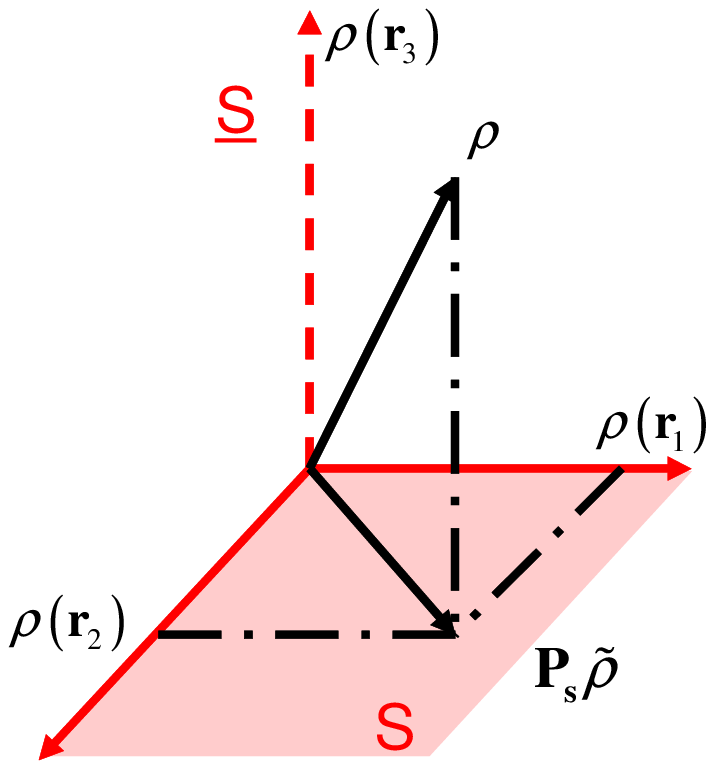}	
    }
\subfigure[]
    {
      \label{subfig:modulus}
	\includegraphics[width=0.2\textwidth]{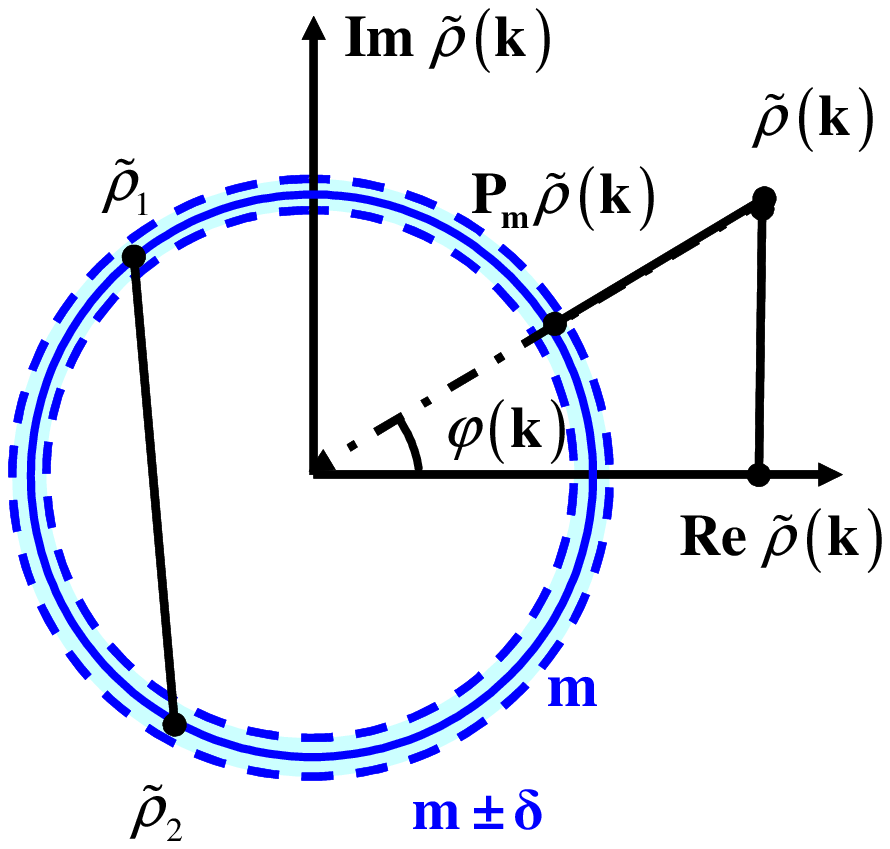}	
    }   
\caption{
Examples of sets and projectors: (a) Support: The axes represent the
values on 3 pixels of an image $\rho$ known to be 0 outside the
support $S$. The vertical axis $\rho(\bm r_3)$ represents a pixel outside
 ($\bm {r}_3 \in \underline{S}$), while the horizontal plane represents pixels inside
 $S$.  The projection on this set is performed simply by
setting to 0 all the pixels outside the support. (b) Modulus: A pixel
(in Fourier space) with a given complex value is projected on the
closest point on the circle defined by the radius $m$. If there is
some uncertainty in the value of the radius $m\pm\delta$, the circle
becomes a band.  The circle is a non-convex set, since the linear
combination between two points on the same set $\rho_1$ and $\rho_2$
does not lie on the set.  Also represented in the figure is the
projection on the real axis (reality projection).
}\label{fig:sets}
\end{figure}

We consider two sets, $S$ (support) and $M$ (modulus).  When the image
belongs to both sets simultaneously, we have reached a
solution.  If the properties of the object being imaged are known
a-priori to be limited in a support region, we know
that in the $n$-dimensional space of the pixel
values, some values must be zero. Images that satisfy this rule
(Eq. (\ref{eq:support})) form the support constraint set.  A projection
onto this set ($\Ps$) involves setting to 0 the components outside
the support, while leaving the rest of the values unchanged
(Fig. \ref{subfig:support}):
\begin{equation}
\bm{P}_s \rho(\bm{r})=
\begin{cases}
\rho(\bm{r}) &\text {if  $\bm{r}\in S$} \\
 \\
0 &\text{otherwise,}
\end{cases}
\end{equation}
and its complementary projector $\Pns={\bm I}-\Ps$. 

The values in every pixel in Fourier space can be described using two
components, the real and imaginary parts, or amplitude and phase,
both defining a point in a complex plane. In an intensity measurement
we obtain the amplitude or {\it modulus} in every pixel that defines a
circle in a complex plane. These circles define the modulus constraint
(Fig. \ref{subfig:modulus}).  When every complex-valued pixel lies on the
 circle defined by the corresponding modulus, the image
satisfies this constraint and it belongs to the {\it modulus} set. Segments
joining two points on a circle do not belong to the circle;
therefore the linear combination of two images is outside the set: the
set is non-convex. These sets are problematic because of the presence
of local minima and undefined projections.

The projection of a point in each complex plane onto the corresponding
circle is accomplished by taking the point on the circle closest to
the current one, setting the modulus to the measured one 
$\sqrt{I(\bm k)}$, and leaving the phase unchanged
(Fig. \ref{subfig:modulus}):
\begin{eqnarray}
\label{eq:modulus}
\tPm \trho(\bm{k})&=&
\tPm |\tilde \rho(\bm{k})|e^{i\varphi(\bm{k})}=\sqrt I
(\bm{k})e^{i\varphi(\bm{k})}\,,
\end{eqnarray}
where we have defined the reciprocal space representation of the projector:
\begin{eqnarray}
\Pm &=&{\cal F}^{-1} \tPm {\cal F}\,,
\end{eqnarray}
and ${\cal F}$ and ${\cal F}^{-1}$ represent the forward and inverse
Fourier transforms respectively.  

This operator is demonstrated to be a projector on the
{\it{non-convex}} (Fig. \ref{subfig:modulus}) set of the magnitude
constraint \cite{luke:siam}.   The same paper discusses the
problems of multi valued projections for non-convex sets, which do not
satisfy the requirements for gradient-based minimization algorithms,
and the related non-smoothness of the squared set distance metric,
which may lead to numerical instabilities.  See also \cite{luke:siam1}
for a follow-up discussion on the non-smooth analysis.

\begin{figure}
\includegraphics[width=0.3\textwidth]{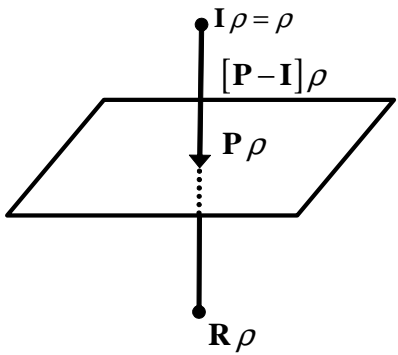}	
\caption{The reflector applies the same step as the projector $(\bm{P-I})$
 twice:
$\bm{R} \rho=\bm{I}\rho +2[\bm{P-I}] \rho$
\label{fig:reflector}
}
\end{figure}

A projector $\bm{P}$ is an operator that takes to the closest point of
a set from the current point $\rho$.  A repetition of the same
projection is equal to one projection alone ($\bm{P^2}=\bm{P}$);
its eigenvalues must therefore be $\lambda=0,1$. Another operator used
here is the reflector $\bm{R}=\bm{I}+2[\bm{P}-\bm{I}]=2\bm{P}-\bm{I}$,
which applies the same step as the projector but moves twice as far 
(Fig \ref{fig:reflector}). In
the case of the support constraint, the whole image space can be
described in terms of the eigenvectors of the corresponding {\it
linear} projector.  These eigenvectors with eigenvalues of 1 (0) are
the images with all the pixels equal to 0, except for one pixel inside
(outside) the support.  The modulus projector is a non-linear
operator:
\begin{eqnarray}
\nonumber
P_m (a+b)&\neq& P_m(a)+P_m(b)\\
P_m (\alpha a)&\neq& \alpha P_m(a)\,,
\end{eqnarray}
and it cannot be described in terms of eigenvalues and eigenvectors.

The Euclidean length $||\rho||$ of a vector $\rho$ is defined as:
\begin{equation}
||\rho||^2=\rho^{\dag} \cdot \rho=\sum_{\bm r} |\rho(r)|^2=
\sum_{\bm k} |\tilde \rho(k)|^2.
\end{equation}
The sum is extended to the measured portion of the diffraction
pattern.  If part of the reciprocal space is not measured, it should
not be included in the sum.  In fact the sum should be weighted with the
experimental noise $\sigma(\bm k)$:
\begin{equation}
||\rho||^2=\frac
{\sum_{\bm k} \tfrac {1}{\sigma^2(\bm k)} |\tilde \rho(k)|^2}
{\sum_{\bm k} \tfrac {1}{\sigma^2(\bm k)} }\,,
\end{equation}
with $\sigma(\bm k)=\infty$ for values of $\bm k$ not measured.
The distance from the current point to the set $||\bm P \rho -\rho||$ is the
basis for our error metric. Typically the errors in real  ($\eps_s$)
and reciprocal space ($\eps_m$) 
are defined in terms of their distance to the corresponding sets: 
\begin{eqnarray}
\nonumber
\eps_s(\rho)&=&\vectornorm{\bm{P}_s \rho-\rho}, \\
\eps_m(\rho)&=&\vectornorm{\bm{P}_m \rho-\rho},
\end{eqnarray}
or their normalized version $\overline
\eps_x(\rho)=\tfrac{\eps_x(\rho)}{\vectornorm{\bm{P}_x \rho}}$.
Another error metric used in the literature is given by the distance
between the two sets: $\eps_{s,m}(\rho)=\vectornorm{\bm{P}_m
\rho-\bm{P}_s \rho}$.  The projector $\bm{P}_m$ moves $\rho$ to the
closest minimum of $\eps_m^2(\bm{P}_m \rho)=0$, providing a simple
relation with the gradient $\nabla_{\rho}\eps^2_m(\rho)$
\cite{Fienup:1982,luke:siam}:
\begin{eqnarray}
\label{eq:gradient}
\bm{P_m}\rho&=&\rho+[\bm{P_m}-\bm{I}]\rho=\rho-
\tfrac 1 2\nabla_{\rho}\eps^2_m(\rho)\,,
\end{eqnarray}
where $\nabla_{\rho}\eps^2_m(\rho)$ is proportional to 
$\nabla_{\rho}\eps_m(\rho)$:
\begin{equation}
\nabla_{\rho}\eps^2_m(\rho)=2\eps_m(\rho)\nabla_{\rho}\eps_m(\rho).
\end{equation}
For $\tilde \rho(\bm k)=0$, $\eps_m$ is non
differentiable, and the projector $\bm{P}_m$ is multivalued
\cite{luke:siam}. The presence of complex zeros ($\rho(\bm k)=0$)
is considered of fundamental importance in the phase-retrieval problem
\cite{fiddy}, and the phase vortices associated with these zeros cause
stagnation in iterative algorithms \cite{fienup:stagnation}.
Several methods have been proposed to solve this problem
\cite{isernia, fienup:stagnation, Marchesini:XRM_CT, Oszlanyi, Miao:2006}.
Similarly the projector $\Ps$ minimizes the error $\eps_s^2$:
\begin{equation}
\label{eq:gradeps_s}
[\bm{I}-\Ps]\rho=\rho_\us=\tfrac 1 2 \nabla_{\rho}\eps_s^2(\rho)
\end{equation}

\section{Iterative projection algorithms}
\label{sec:IPA}

Several algorithms based on these concepts have now been proposed and
a visual representation of their behavior is useful to characterize
the algorithm in various situations, in order to help choose the most
appropriate one for a particular problem. In this section the
projection algorithms published in the literature are summarized (see
also Table \ref{tab:algorithms}) and tested on simple geometrical sets.

\begin{table}
\caption{\label{tab:algorithms}
Summary of various algorithms}
\begin{ruledtabular}
\begin{tabular}{|l|l|}
Algorithm & Iteration $\rho^{(n+1)}=$\\
\hline
\hline
ER & $\bm{P_s P_m}\rho^{(n)}$\\
\hline
SF &$\bm{R_s P_m}\rho^{(n)}$\\
\hline
HIO &\label{eq:HIOtab}
$
\begin{cases}
\bm{P_m} \rho^{(n)}(\bm{r})  & 
	\text {$\bm{r}\in S$} \\
(\bm{I}-\beta \bm{P_m})\rho^{(n)}(\bm r)  & \bm{r}\notin S
\end{cases}$ \\
\hline
DM &
$
\begin{array}{llll}
\{
\bm{I}&+&\beta \bm{P_s} &
 \left [
  \left (
	1+\gamma_s
  \right )
  \bm{P_m}-\gamma_s\bm{I}
 \right ]\\
&-&\beta \bm{P_m} &
 \left [
  \left (
	1+\gamma_m
  \right )
  \bm{P_s}-\gamma_m \bm{I}
 \right ]
 \}
\rho^{(n)} 
\end{array}$\\
\hline
ASR&$\tfrac{1}{2}[\bm{R_s R_m}+\bm{I} ]\rho^{(n)}$\\ 
\hline
HPR&$\tfrac{1}{2}[
	\bm{R_s} 
	\left (
		\bm{R_m}+(\beta-1) \bm{P_m} 
	\right )$\\
&$
	+\bm{I}
	+(1-\beta )\bm{P_m}
	]
\rho^{(n)} 
$\\
\hline
RAAR&$
\left [ \tfrac{1}{2} \beta \left (
	\bm{R_s R_m}+\bm{I} 
	\right )
	+(1-\beta)\bm{P_m}
	\right ] 
\rho^{(n)} $\\
\end{tabular}
\end{ruledtabular}
\end{table}

\begin{figure}
\subfigure[ER]
    {
	\includegraphics[width=0.2\textwidth]{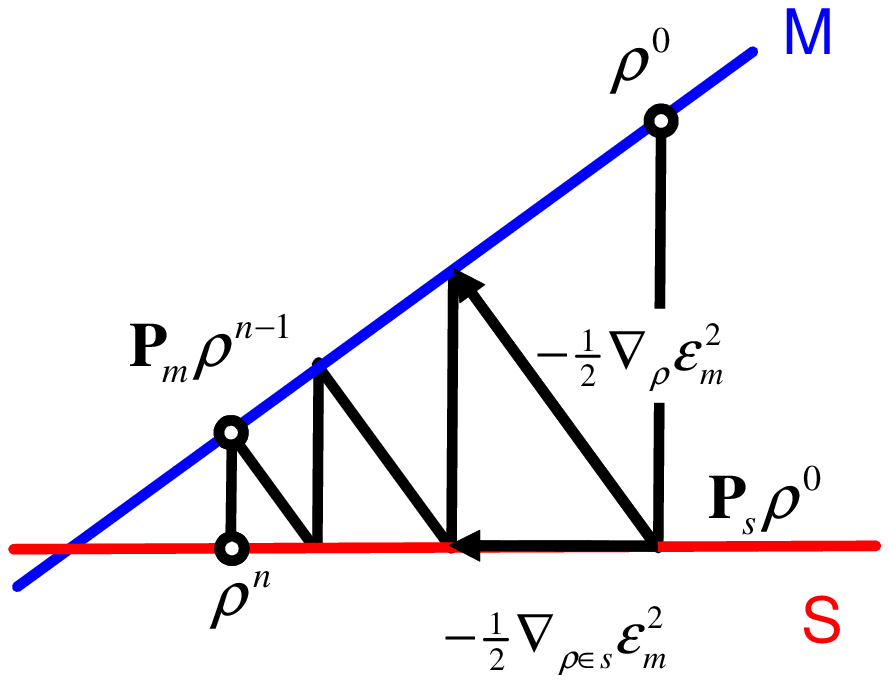}	
      \label{subfig:ER}
    }
\subfigure[Solvent Flip]
    {
	\includegraphics[width=0.2\textwidth]{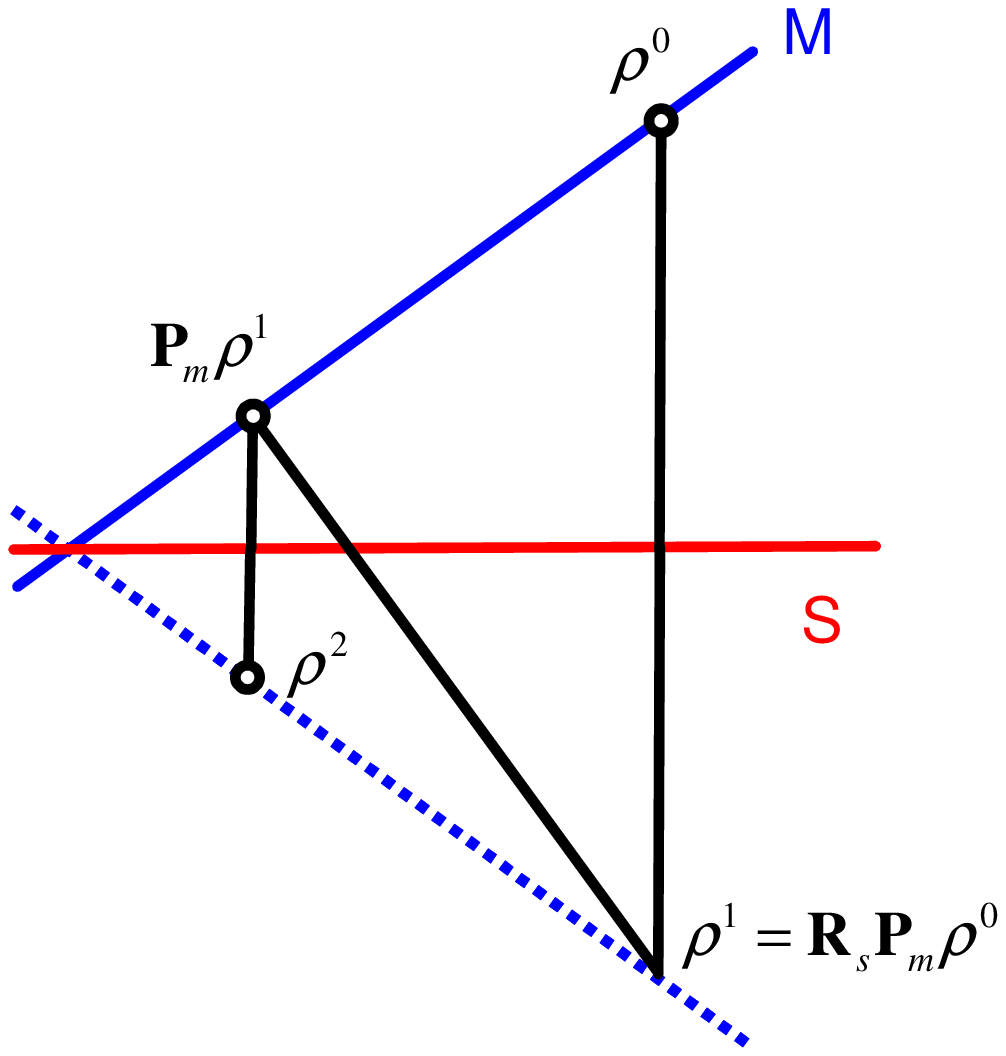}	
      \label{subfig:chargeflip}
    }\\
\subfigure[HIO]
    {
      \label{subfig:HIO}
	\includegraphics[width=0.22\textwidth]{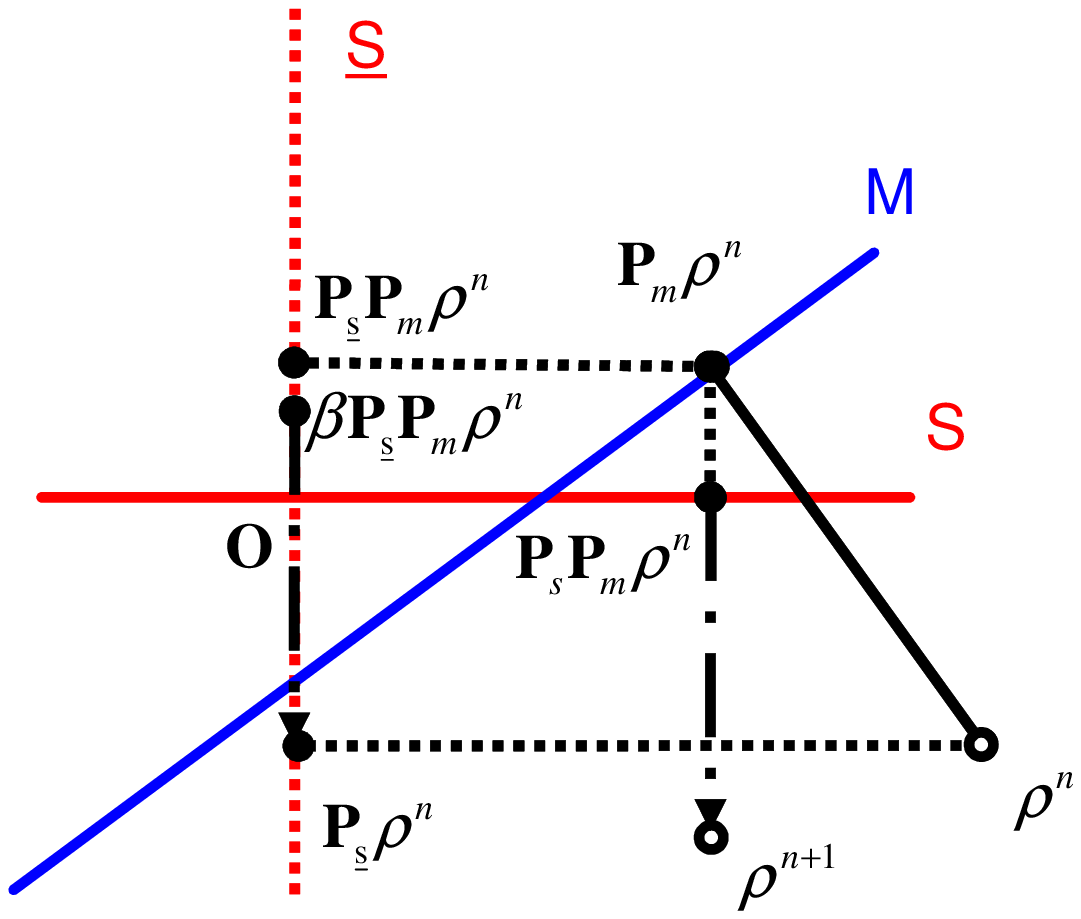}	
    }
\subfigure[Difference Map]
    {
      \label{subfig:DM}
	\includegraphics[width=0.22\textwidth]{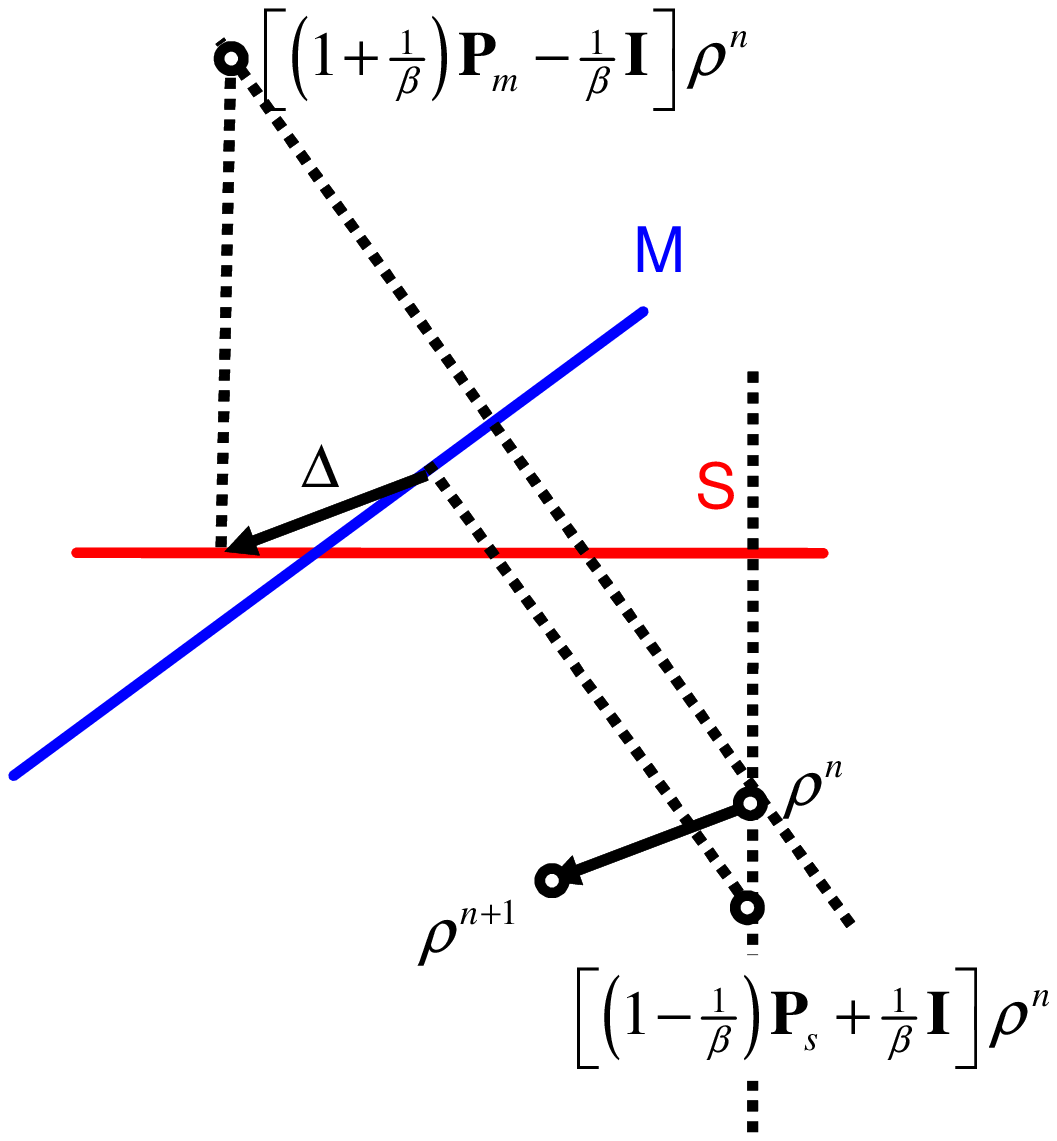}	
    }
\caption{Geometric representation of various algorithms using a simplified version of the constraint: two lines intersecting. (a) Error reduction algorithm: we start from a point on the 
modulus constraint by assigning a random phase to the diffraction pattern.
The projection onto the modulus constraint finds the point on the set 
which is nearest to the current one. The arrows indicate the gradients 
of the error metric. (b) The speed of convergence is increased by replacing the
projector on the support with the reflector. The algorithm jumps between the 
modulus constraint (solid diagonal line) and its mirror image with respect to the support constraint (dotted line).
(c) Hybrid input--output, see text (Eq. (\ref{eq:HIO})). The space perpendicular to the support set is represented by the vertical dotted line $\underline{S}$. 
(d) Difference map, see text (Eq. (\ref{eq:DM})).
\label{fig:algorithms}
}
\end{figure}

The following algorithms require a starting point $\rho^0$, which is
generated by assigning a random phase to the measured object amplitude
(modulus) in the Fourier domain $|\tilde
\rho(\bm{k})|=m(\bm{k})=\sqrt{I(\bm{k})}$.  The first algorithm called
{\it error reduction} (ER) (Gerchberg and Saxton
\cite{Gerchberg:1972, bregman:1965, stark:1984}) is simply (Fig. \ref{subfig:ER}):
\begin{equation}
\rho^{(n+1)}=\bm{P_s P_m}\rho^{(n)}\,,
\end{equation}
and by projecting back and forth between two sets, it converges to the
local minimum. The name of the algorithm is due to the steps moving
along the gradient of the error metric (see Eq. (\ref{eq:gradient})):
\begin{equation}
\label{eq:ER}
\Ps\Pm\rho=\Ps \rho-
\tfrac 1 2\nabla_{s}\,\eps^2_m(\rho)
\,,
\end{equation}
where $\nabla_{s}=\Ps \nabla$ is the component of the
gradient in the support. Figure \ref{subfig:ER} shows that the step size
is far from optimum, but that it guarantees linear convergence. A line search
along this gradient direction would considerably speed up the
convergence to a local minimum and will be discussed in Section 
\ref{sec:gradients}.

The {\it solvent flipping} (SF) algorithm \cite{abrahams:1996} is obtained
by replacing the support projector
$\bm{P}_s$ with its reflector $\bm{R}_s=2\bm{P}_s-\bm{I}$ (Fig. \ref{subfig:chargeflip}) :
\begin{equation}
\rho^{(n+1)}=\bm{R_s P_m}\rho^{(n)} \,,
\end{equation}
\noindent
which multiplies the charge density $\rho$ outside the support by $-1$.
The {\it hybrid input-output} (HIO) \cite{Fienup:1978,Fienup:1982}
(Fig. \ref{subfig:HIO}) is based on non-linear feedback control theory
and can be expressed as:
\begin{equation}
\label{eq:HIO}
 \rho^{(n+1)}(x)=
\begin{cases}
\bm{P_m} \rho^{(n)}(x)  & 
	\text {if  $x\in S$,} \\
(\bm{I}-\beta \bm{P_m})\rho^{(n)}(x)  & \text{otherwise.}
\end{cases}
\end{equation}
 Equations (\ref{eq:gradient}) and (\ref{eq:gradeps_s}) can
be used to describe the steps ($\Delta \rho=\rho^{(n+1)}-\rho^{(n)}$) in 
terms of the gradients of the error
metrics. In Section \ref{sec:gradients} it will be shown that this 
algorithm seeks the saddle point:
\begin{equation}
\min_{\rho_s} \max_{\rho_\us} {\cal L} (\rho),\,\,\,\,\,{\cal L}(\rho)= 
 \eps^2_m(\rho)-\eps^2_s(\rho)\,
\end{equation}
by moving in the descent--ascent direction ($[-\Ps+\beta \Pns ]\nabla{\cal L}$) (see
Section \ref{sec:gradients} for details), rather than in the simple error-minimization direction. 

It is often used in conjunction with the ER algorithm, alternating
several HIO and one or more ER iterations (HIO(20)+ER(1) in our
case).  In particular one or more ER steps are used at the end of the
iteration.  Elser \cite{elser:2003} pointed out that the \textit{iterate}
$\rho^{n}$ can converge to a \textit{fixed point}
($\rho^{n+1}=\rho^n$), which may differ from the solution $\bar \rho$
($\bm P_s \bar \rho=\bm{P}_ m \bar \rho=\bar \rho$). 
However the solution $\bar \rho$ can be easily obtained from the fixed point:
\begin{eqnarray}
\label{eq:fixed_point}
\bar \rho^n_m&=&\bm{P}_m \rho^n\,, \\
\bar \rho^n_s&=&(1+\tfrac 1 \beta) \bm{P}_s \bm{P}_m \rho^n - \tfrac 1 \beta \bm{P}_s \rho^n\,,
\nonumber
\end{eqnarray}
where $\bar \rho_m$ and $\bar \rho_s$ should coincide, or else their
difference can be used as an error metric. See \cite{elser:2003} for
further details.

\begin{figure*}
\subfigure[]
    {
      \label{subfig:2lines2d}
	\includegraphics[width=0.4\textwidth]{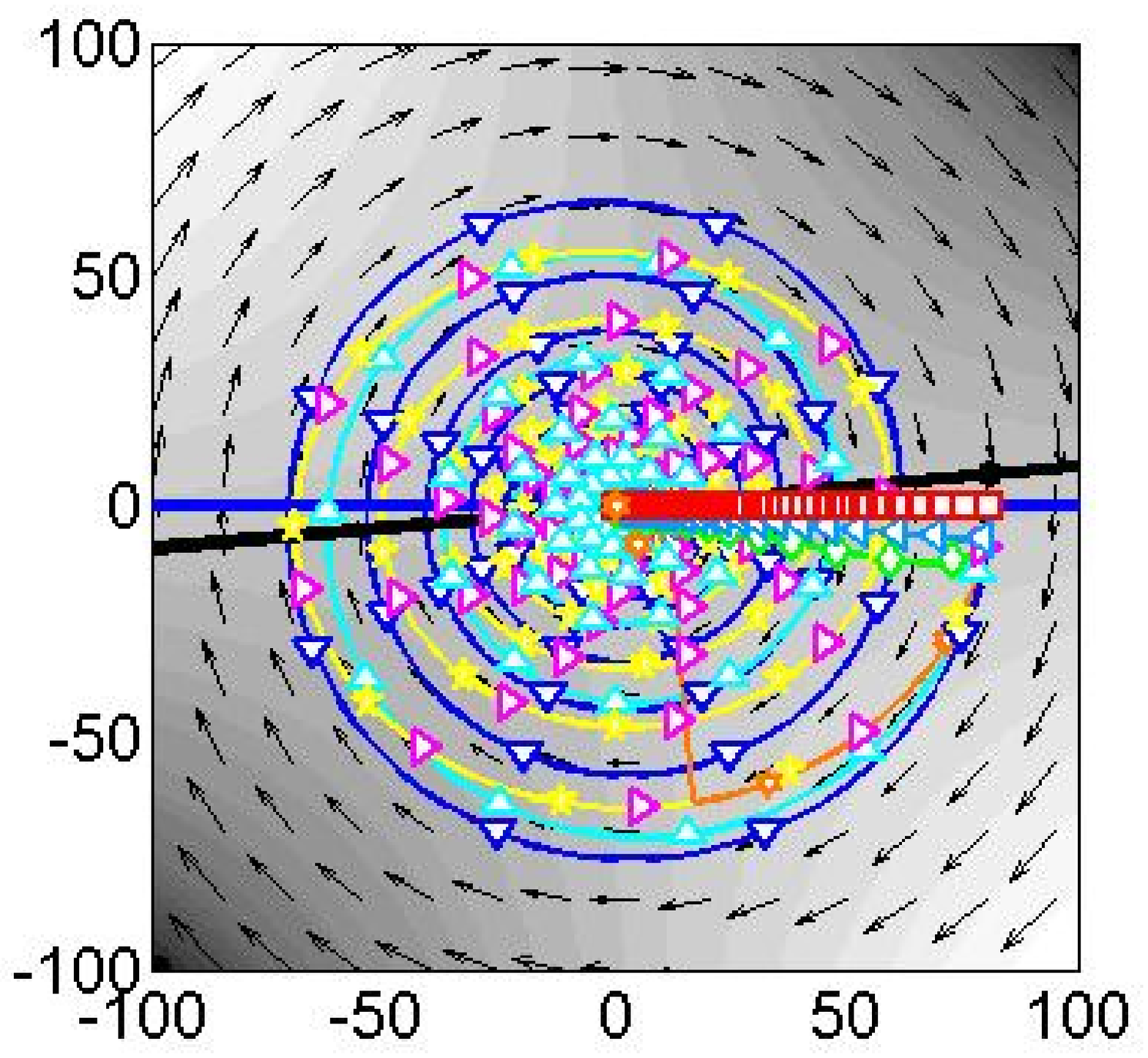}
	\usebox{\mylegend}
    }
\subfigure[]
    {
      \label{subfig:2lines3d}
		\includegraphics[width=0.4\textwidth]{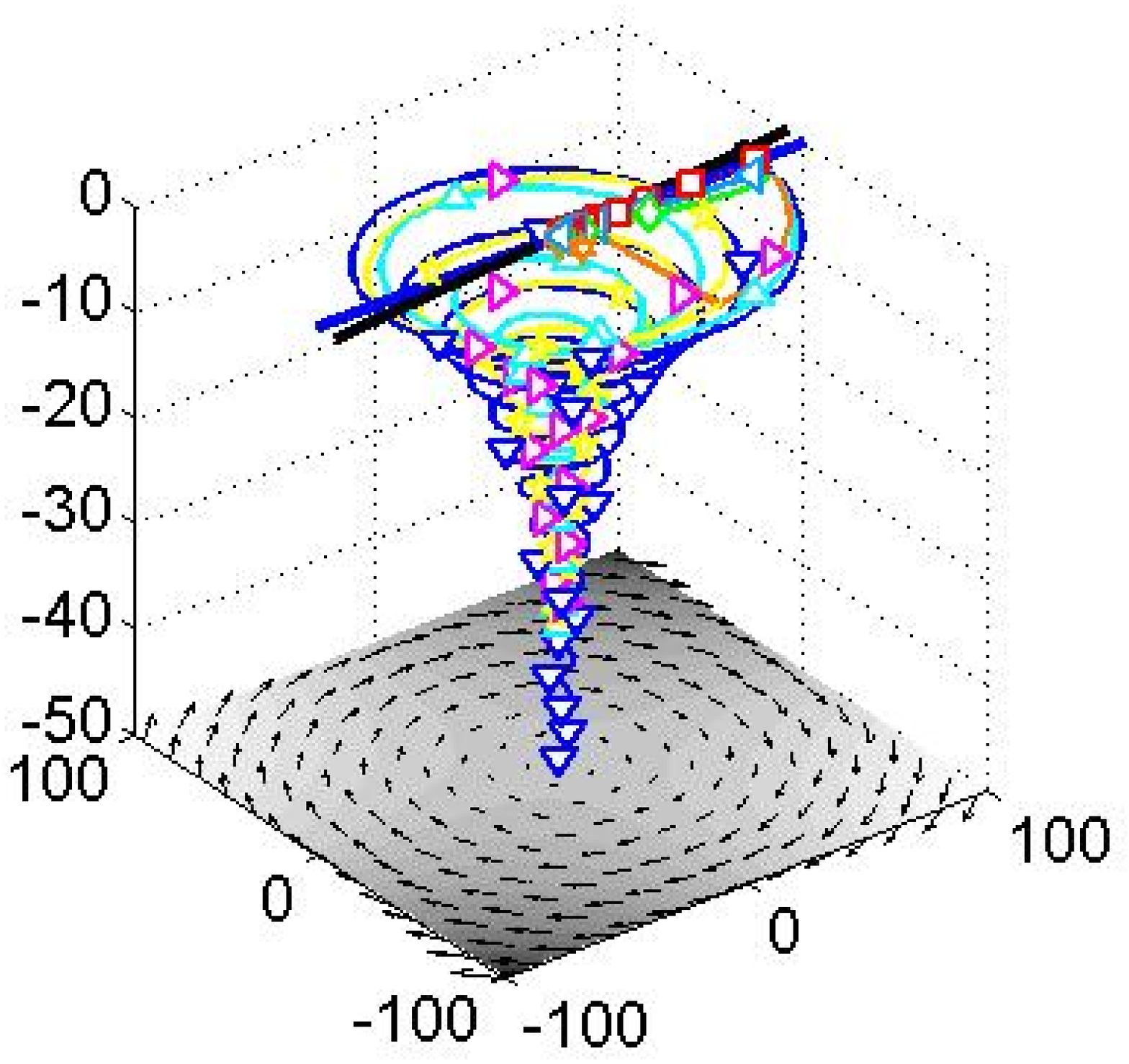}
    }
\caption{
The basic features of the iterative projection algorithms can
be understood by this simple model of two lines intersecting
(a).  The aim is to find the intersection.  The ER
algorithm and the solvent flipping algorithms converge in some
gradient-type fashion (the distance to the two sets never increases),
the solvent flip method being slightly faster when the angle
between the two lines is small.  HIO and variants move following a spiral path.
The lagrangian (${\cal L}=\eps^2_m-\eps^2_s$) is represented in grayscale,
and the descent-ascent directions ($[-\nabla_s,\nabla_\us ]{\cal L}$)
are indicated by arrows.  When the two lines do not
intersect (b), HIO and variants keep moving in the
direction of the gap between the two lines, away from the local minimum.
 ER, SF and RAAR converge at (or close to) the local minimum.
\label{2lines}
}
\end{figure*}

The {\it difference map} (DM) is a general set of algorithms 
\cite{elser:2003},  which requires 4 projections (two
time-consuming modulus constraint projections) (Fig. \ref{subfig:DM}):
\begin{eqnarray}
\nonumber
\rho^{(n+1)}=
\{&\bm{I}&+\beta \bm{P_s} 
 \left [
  \left (
	1+\gamma_s
  \right )
  \bm{P_m}-\gamma_s\bm{I}
 \right ]\\
&-& \beta \bm{P_m} 
 \left [
  \left (
	1+\gamma_m
  \right )
  \bm{P_s}-\gamma_m \bm{I}
 \right ]
\}
\rho^{(n)} \,;
\label{eq:DM}
\end{eqnarray}
the solution corresponding to the fixed point is described in the same
article \cite{elser:2003}. We will use in the upcoming tests what Elser
 suggested as the optimum, with $\gamma_s=-\beta^{-1}$ and $\gamma_m=\beta^{-1}$.

The {\it averaged successive reflections} (ASR) \cite{luke:1} algorithm is:
\begin{equation}
\rho^{(n+1)}=\tfrac{1}{2}[\bm{R_s R_m}+\bm{I} ]\rho^{(n)} \,.
\end{equation}
The {\it Hybrid Projection Reflection} (HPR) \cite{luke:2} algorithm
 is derived from a relaxation of the ASR:
\begin{eqnarray}
\nonumber
\rho^{(n+1)}&=&\tfrac{1}{2}[
	\bm{R_s} 
	\left (
		\bm{R_m}+(\beta-1) \bm{P_m} 
	\right )\\
	&+&\bm{I}
	+(1-\beta )\bm{P_m}
	]
\rho^{(n)}\,. 
\end{eqnarray}
It is equivalent to HIO if positivity (Section \ref{sec:positivity}) is
not enforced, but it is written in a recursive form, instead of a case-by-case 
form such as Eq. (\ref{eq:HIO}). It is also equivalent to the
DM algorithm for $\gamma_s=-1$, $\gamma_m=\beta^{-1}$.  Finally the {\it
relaxed averaged alternating reflectors} (RAAR) algorithm
\cite{luke:3}:
\begin{equation}
\rho^{(n+1)}=\left [ \tfrac{1}{2} \beta \left (
	\bm{R_s R_m}+\bm{I} 
	\right )
	+(1-\beta)\bm{P_m}
	\right ] 
\rho^{(n)}\,. 
\end{equation}
For $\beta=1$, HIO, HPR, ASR and RAAR coincide.

The first test is performed on the simplest possible case: find the
intersection between two lines. Figure \ref{2lines} shows the behavior
of the various algorithms. The two sets are represented by a
horizontal blue line (support) and a tilted black line (modulus).  ER
simply projects back and forth between these two lines, and moves
along the support line in the direction of the intersection.  SF 
projects onto the modulus, `reflects' on the support, and moves
along the reflection of the modulus constraint onto the support.  The
solvent flipping algorithm is slightly faster than ER thanks to the
increased in the angle between projections and reflections.  HIO and
variants (ASR, DM, HPR) move in a spiral around
the intersection, eventually reaching the intersection. For similar
$\beta$ RAAR behaves somewhere in between ER and HIO with a sharper
spiral, reaching the solution much earlier. Alternating 20 iterations
of HIO and 1 of ER (HIO(20)+ER(1)) considerably speeds up the convergence.

When a gap is introduced between the two lines (Fig. \ref{subfig:2lines3d}),
 so that they do not intersect, HIO and variants move
away from this local minimum in search of another `attractor' or local
minimum. This shows how these algorithms escape from local minima and explore
the multidimensional space for other minima. ER, SF and 
RAAR converge to or near the local minimum. 
By varying $\beta$ RAAR becomes a local minimizer for small $\beta$, and
becomes like HIO for $\beta\simeq 1$. ER, SF and HIO+ER converge
to the local minimum in these tests. 

\begin{figure}
\includegraphics[width=0.32\textwidth]{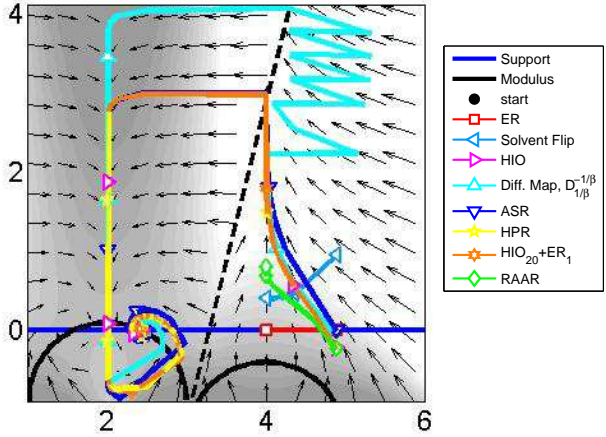}
	\usebox{\mylegend}
\caption{The horizontal line represents a support constraint, 
while the two circles represent a non-convex constraint, i.e.  the
modulus constraint.  The gradient-type (ER and SF) algorithms converge to
the local minimum, while HIO and variants follow the descent-ascent direction ($[-\nabla_s,\nabla_\us ]{\cal L}$)  indicated by the 
arrows.\label{fig:2circles}
}
\end{figure}

A more realistic example is shown in Fig. \ref{fig:2circles}. 
Here the circumference of two circles represents a
non-convex set  (modulus constraint), while the support
constraint is represented by a line. 
The convex set represents a simplified modulus constraint in
 a phase-retrieval problem. The advantage of this example
is the simplicity in the `modulus' projector operator (it projects onto 
the closest circle). 

We start from a position near the local minimum. ER, SF, and
HIO+ER  fall into this trap (Fig. \ref{fig:2circles}).  HIO and variants
 move away from the local minimum, `find' the other
circle, and converge to the center of the circle. In the center the
projection on the modulus constraint becomes `multivalued', and its
distance metric is `non-smooth'. Such a point is unstable,
and the algorithms start spiralling toward the solution.
 For $\beta=0.75$, RAAR does not reach the solution, but converge close to the local minimum.

\begin{figure}
\begin{center}
\includegraphics[width=0.32\textwidth]{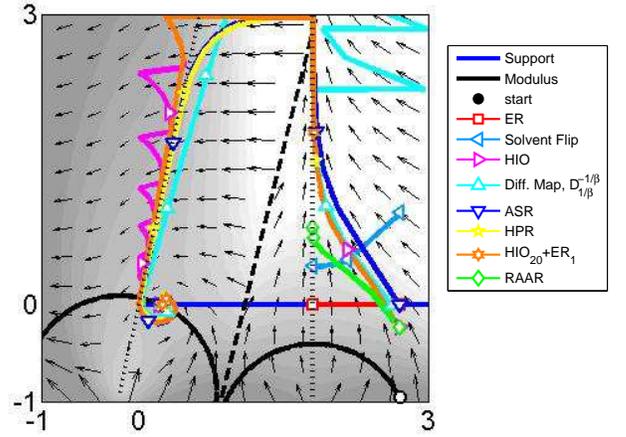}
\usebox{\mylegend}
\end{center}
\caption{
Positivity constraint: the support constraint is represented by a
horizontal line originating from 0 ($x\geq 0$).   A barrier due to the positivity constraint 
changes the behavior of the algorithms, which no longer follow the descent--ascent 
direction. HIO bounces on the $x=0$ axis, while the other algorithms are smoother.}
\label{fig:2circlespositivity}
\end{figure}

\subsection*{Positivity}
\label{sec:positivity}
The situation changes slightly when we consider the positivity constraint. 
The previous definitions of the algorithms still apply, just replacing 
$\Ps$ with $\bm{P}_{s+}$:
\begin{equation}
\bm{P}_{s+}\rho{\bm r}=\begin{cases}
 \rho(\bm r)\ &\text{if $\bm r \in S$ \& $\rho(\bm r)\ge 0$} \,,\\
0 &\text{otherwise.}
\end{cases}
\end{equation}
 The only difference is for HIO which becomes:
\begin{equation}
 \rho^{(n+1)}=
\begin{cases}
\bm{P_m} \rho^{(n)}(\bm r )  & 
	\text {if  $\bm r \in S$ \& $\bm{P_m} \rho^{(n)}(\bm r)\ge 0$}, \\
(1-\beta \bm{P_m})\rho^{(n)}  & \text{otherwise.}
\end{cases}
\end{equation}
HIO and variants follow the saddle-point direction, moving away
from local minima (Fig. \ref{fig:2circlespositivity}), but as they approach the
solution they react differently to the positivity constraint, with HIO
"bounching" at the $x=0$ axis, and ASR/HPR/DM proceeding more smoothly toward 
the solution. 
\section{Steepest descent, 
Conjugate gradient, and min-max algorithms 
\label{sec:gradients}}

\begin{figure*}
\subfigure[Reduced Gradient Method (ER)]
{\label{G:ER}
\includegraphics[width=0.3\textwidth]{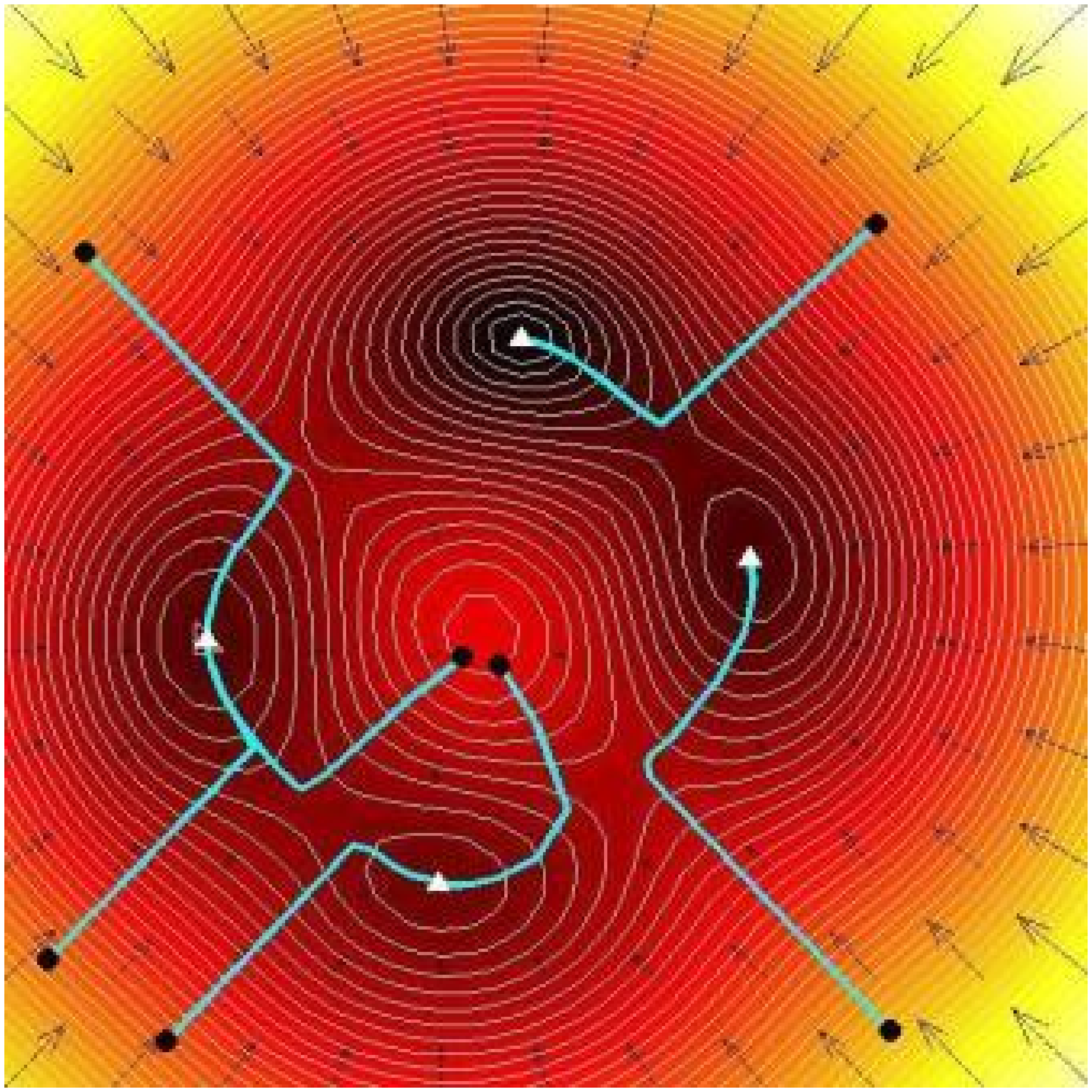}
}
\subfigure[Steepest Descent]
{\label{G:SD}
\includegraphics[width=0.3\textwidth]{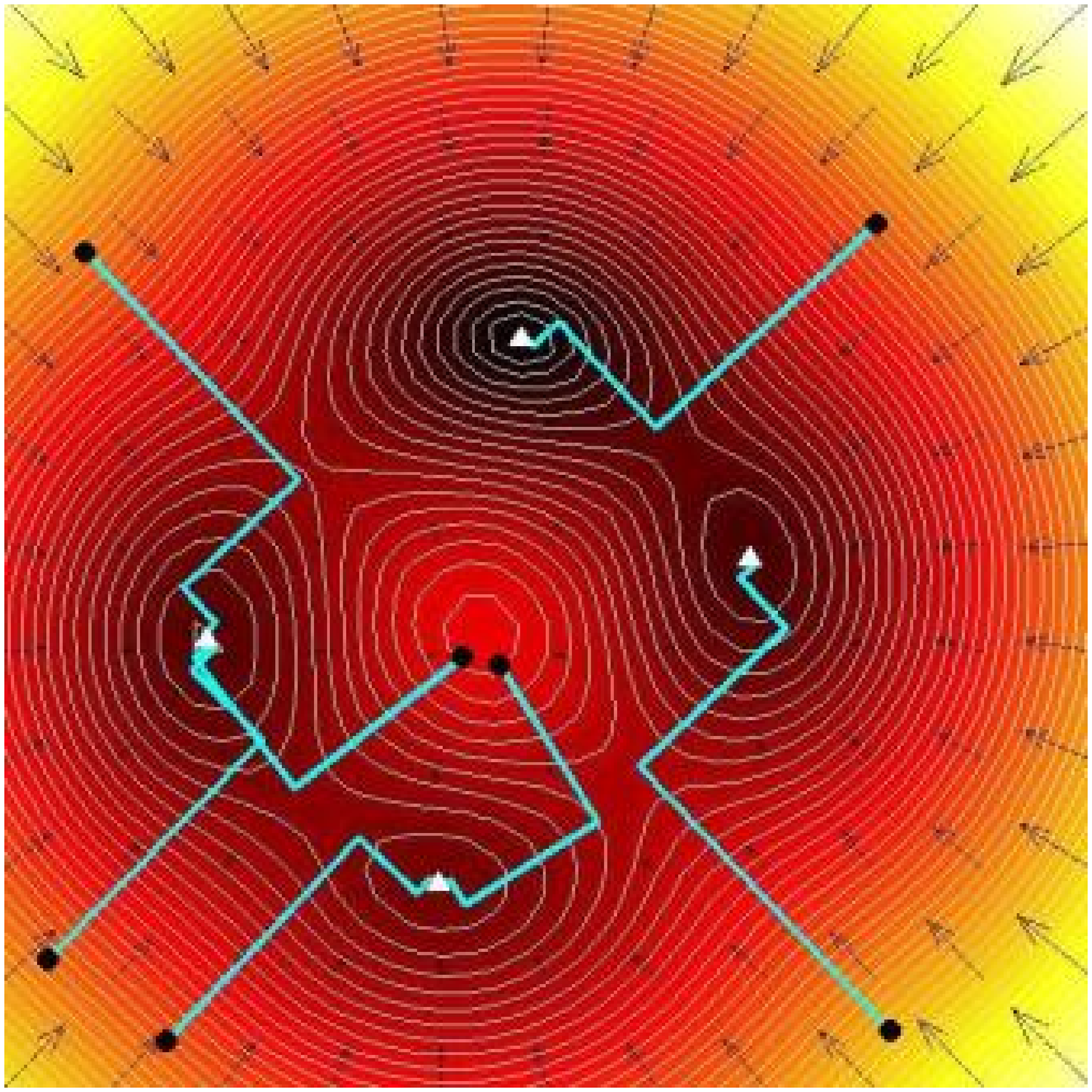}
}
\subfigure[Conjugate Gradient]
{\label{G:CG}
\includegraphics[width=0.3\textwidth]{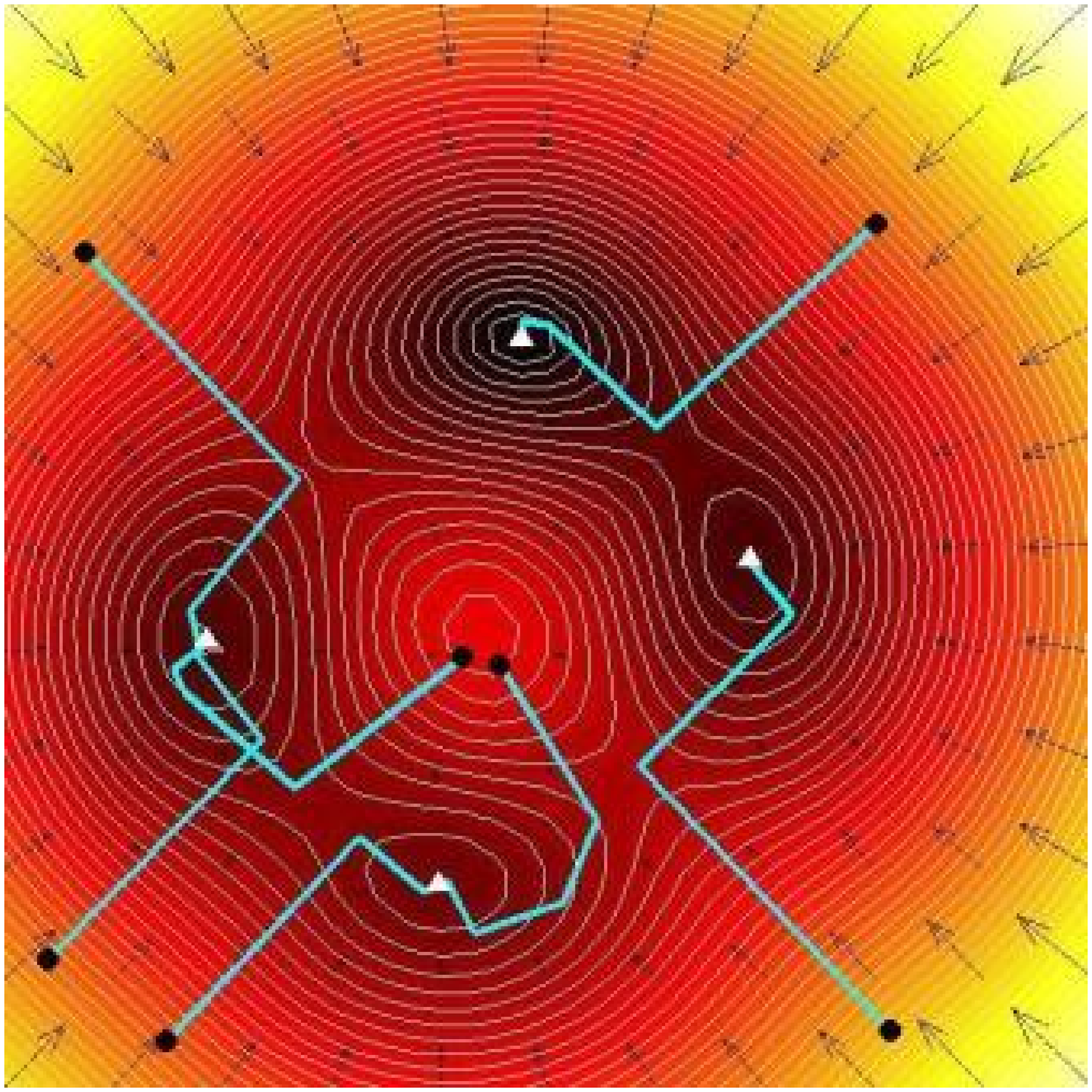}
}
\\
\subfigure[Reduced gradient saddle optimization (HIO/ASR)]
{\label{G:HIO}
\includegraphics[width=0.3\textwidth]{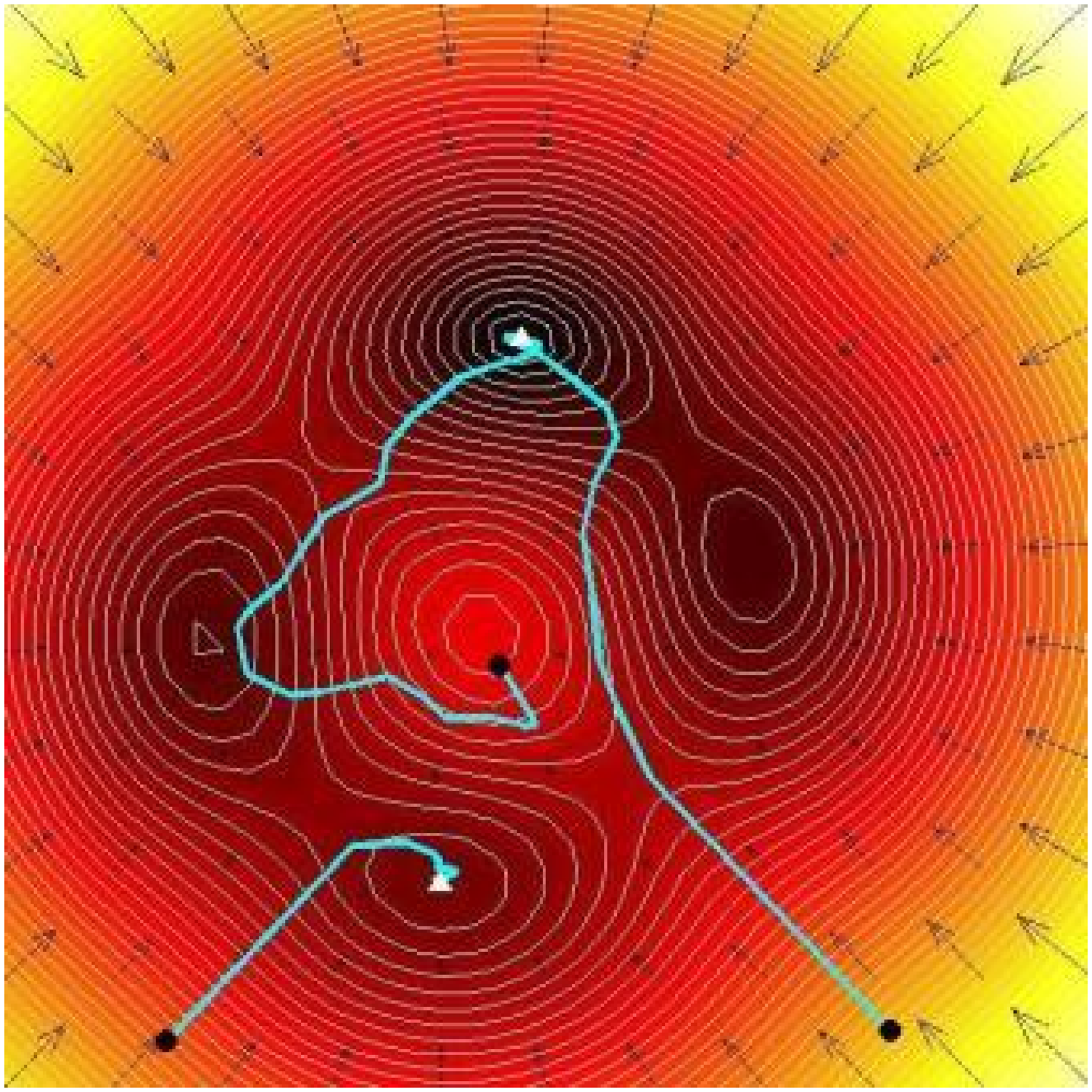}
}
\subfigure[1D saddle optimization]
{\label{G:HIOSD}
\includegraphics[width=0.3\textwidth]{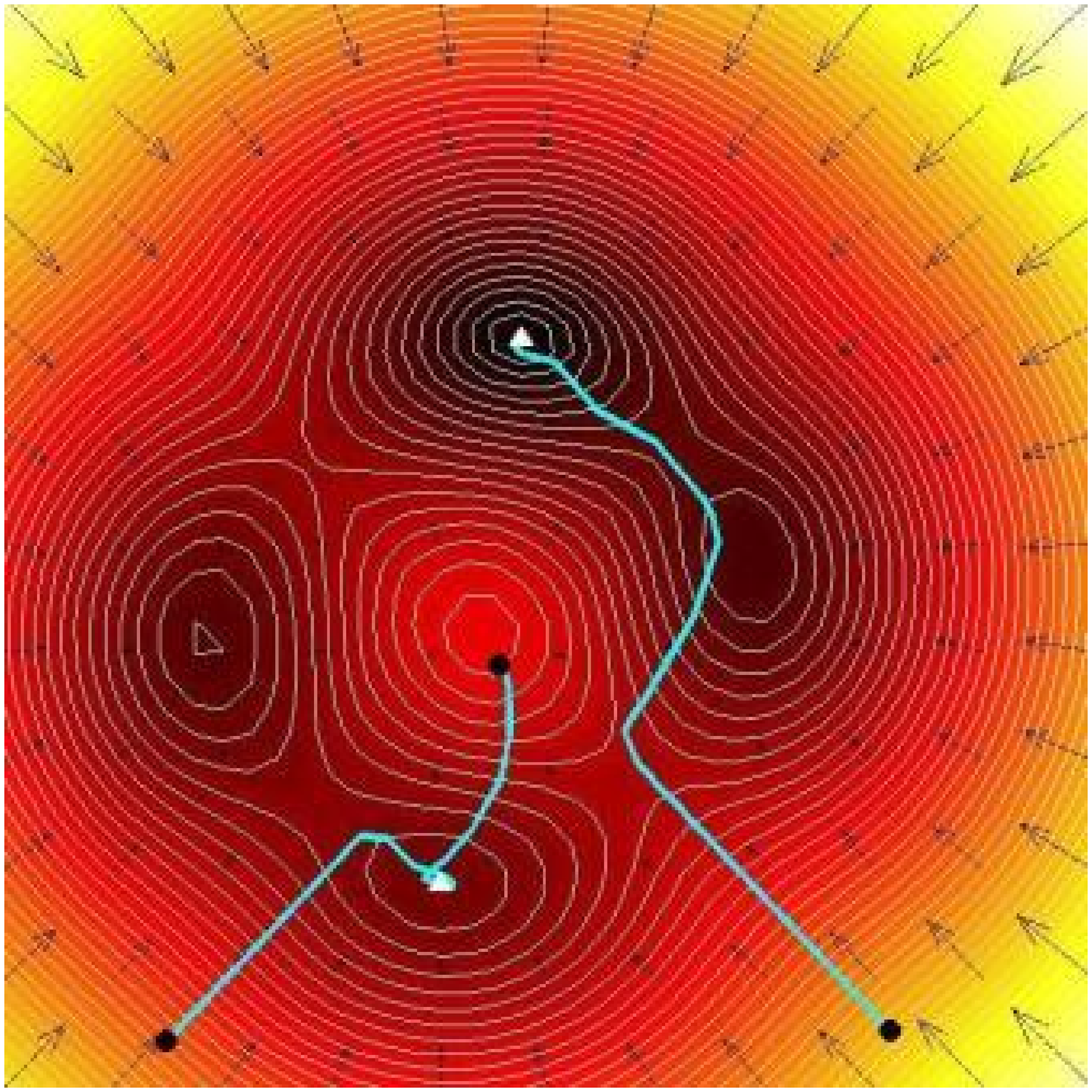}
}
\subfigure[Conjugate saddle optimization]
{\label{G:HIOCG}
\includegraphics[width=0.3\textwidth]{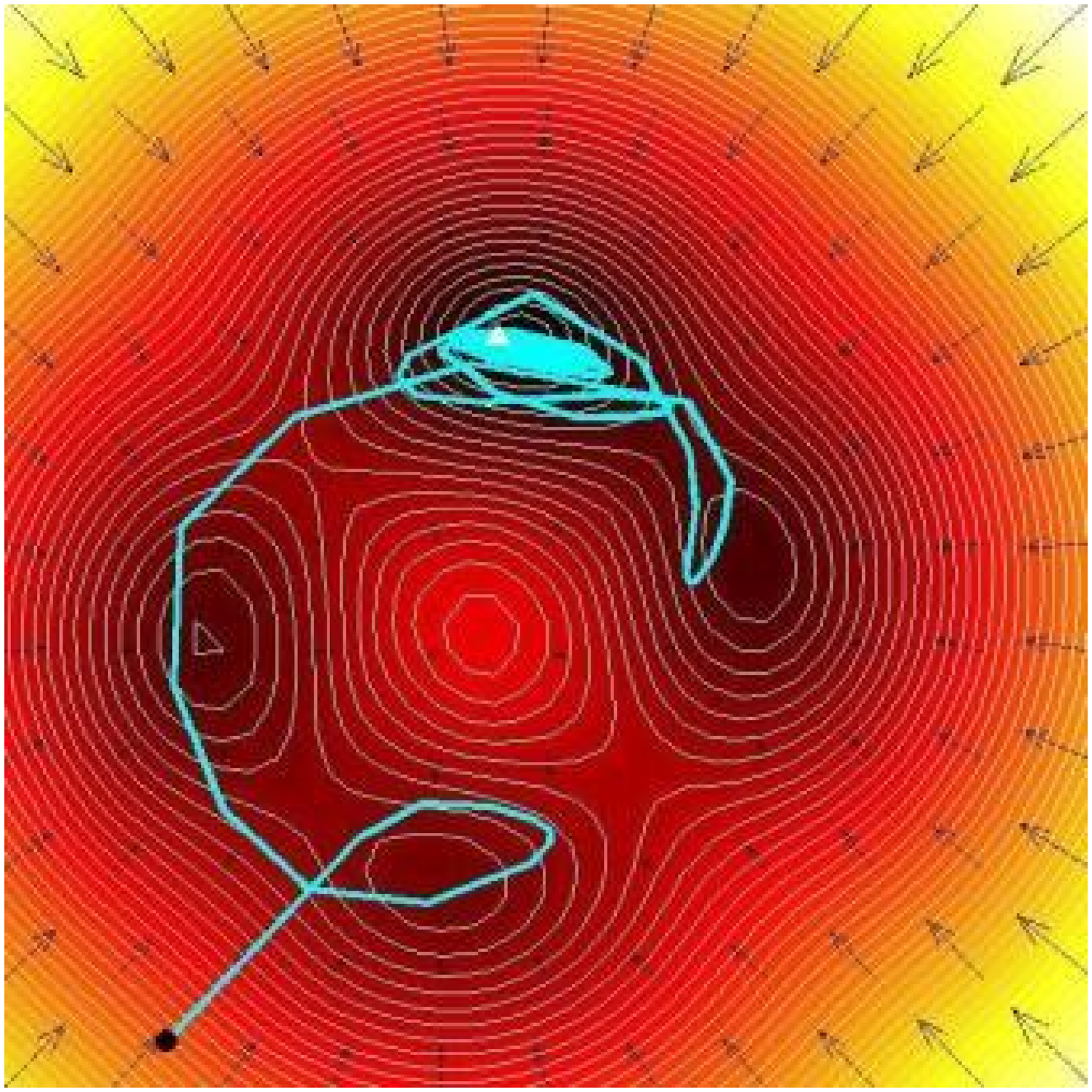}
}\\[-0pt]
\subfigure[2D saddle optimization]
{\label{G:HIOSD2}
\includegraphics[width=0.3\textwidth]{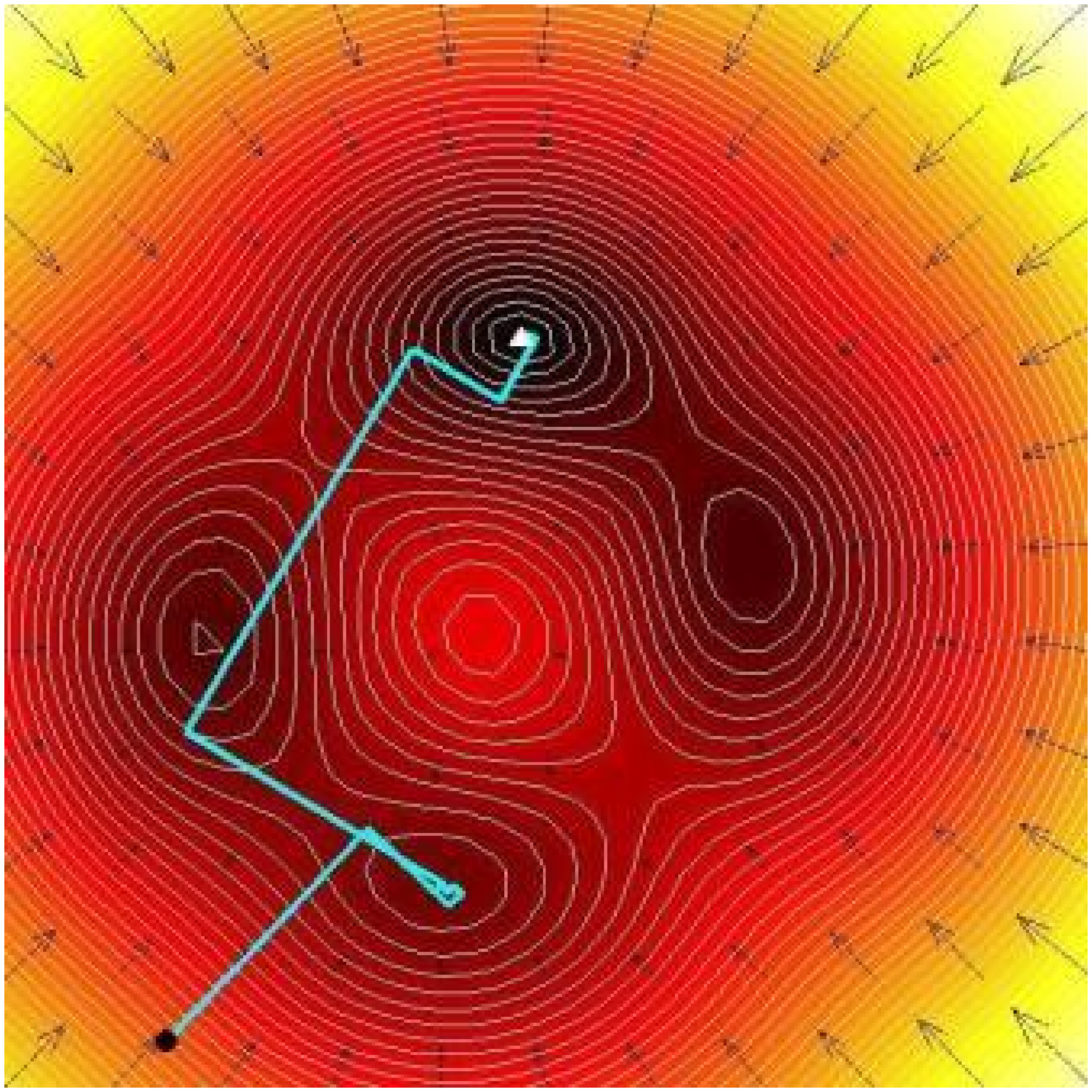}
}
\subfigure[2D conjugate saddle optimization (1)]
{\label{G:HIOCG2}
\includegraphics[width=0.3\textwidth]{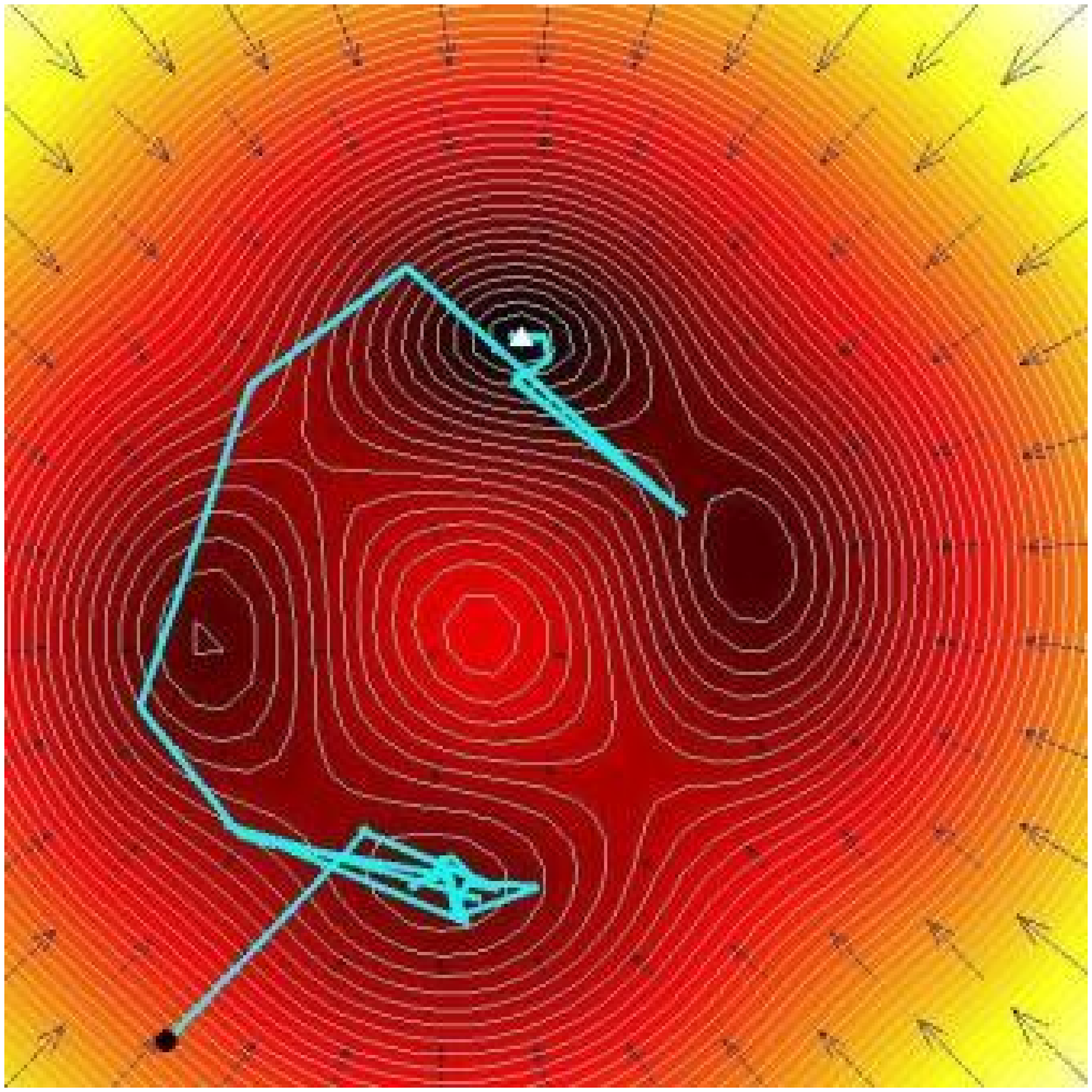}
}
\subfigure[2D conjugate saddle optimization (2)]
{\label{G:HIOCG2s}
\includegraphics[width=0.3\textwidth]{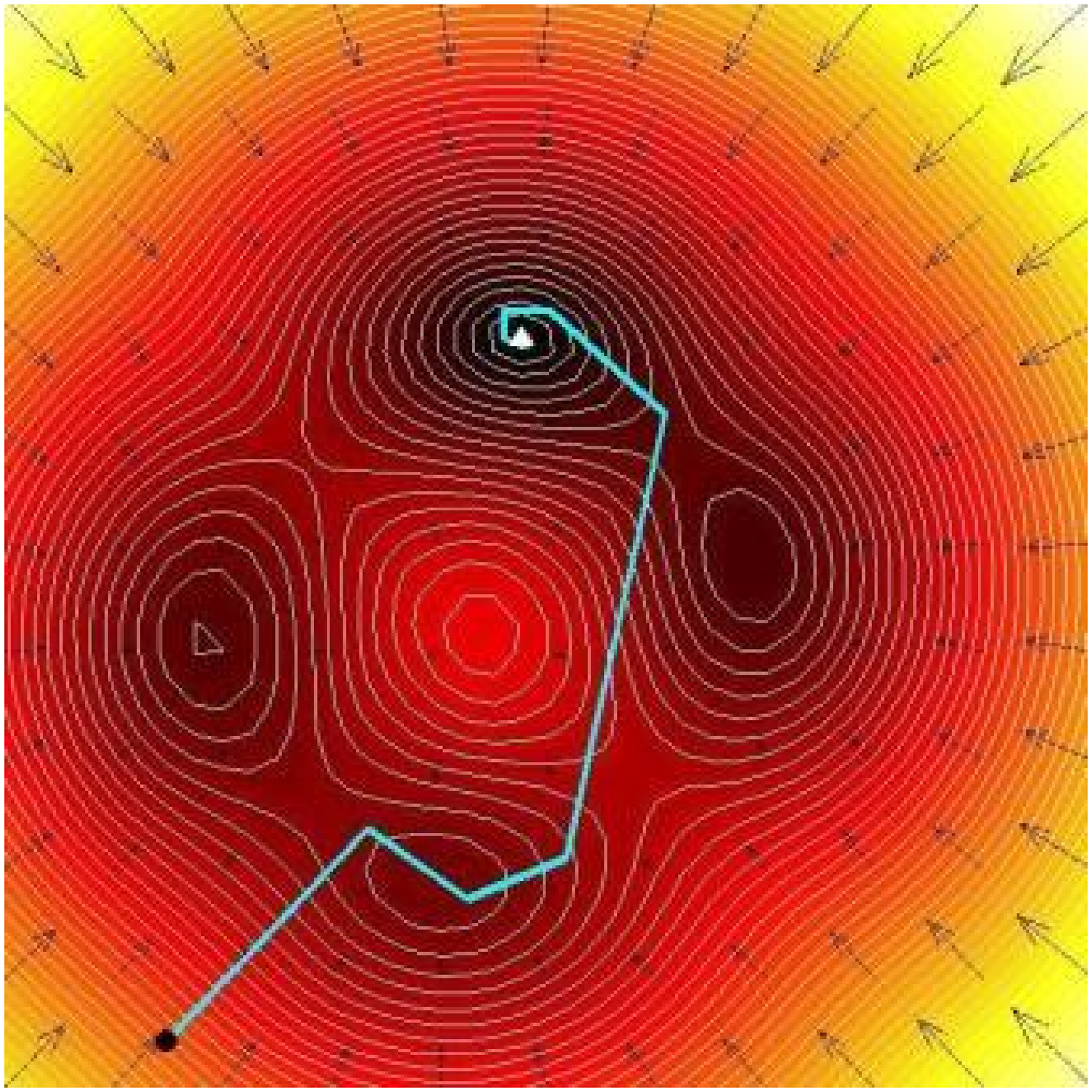}
}
\caption{
\label{fig:gradient}
A simple 2-D phase-retrieval problem: only two variables (pixel values) are unknown.
 The solution --the global minimum-- is the
top minimum in the figures. The colormap and contour lines represents
the error metric $\eps_m(\rho_s)$, and the descent direction is
indicated by the arrows.  The  error reduction algorithm (a) proceeds toward
the local minimum without optimizing the step length and stagnates at
the local minima. The steepest descent method (b) moves toward the local
minimum with a zig-zag trajectory, while the conjugate gradient method
reaches the solution faster (c). The HIO method generally converges to the global minimum, 
however some rare starting points converge to a local minimum (d).  The saddle-point optimization 
with optimized step length (Eq.  {\ref{eq:saddle1d}})
 stagnates in the same local minimum as HIO (e). The conjugate
gradient version avoids stagnation (f). The saddle point optimization
using a two dimensional search of the saddle point reaches the global minimum
from a larger range of starting points than HIO (g). The conjugate gradient
version (h, i) reaches the solution faster if the conjugate directions $\Lambda \trho_{s,\us}$ are
obtained independently from $\Delta \trho_{s,\us}$ (i), rather than their sum $\Delta \trho=\Delta \trho_s+\Delta \trho_\us$.  }
\end{figure*}

As discussed in Section \ref{sec:IPA}, the error reduction (ER) algorithm
moves in the direction of the steepest descent \cite{Fienup:1982};
however the step length is not optimized to reach the local minimum in
that direction, since it is only one component of the full gradient
(Fig. \ref{subfig:ER}). Such strategy is generally referred to as reduced
gradient method. 
 Figure \ref{fig:gradient} shows the error metric
$\eps_m$ as a function of two unknown pixel
values in a simple two-dimensional phase-retrieval problem,
 and the behavior of the ER algorithm toward the local minima.

The simplest acceleration strategy, the steepest descent method,
uses the steepest direction (gradient) and performs a line search of
the local minimum in the descent direction:
\begin{eqnarray}
 &\min_{\delta} \eps_m^2 \left (\rho+\delta \Delta \rho \right)\,,&\\
\nonumber
&\Delta \rho=-\half \nabla_{s} \eps^2_m(\rho)
=-\Ps [\bm I-\Pm] \rho\,,&
\end{eqnarray}
where $\nabla_{s}=\bm P_s \nabla_{\rho}$ is the gradient with respect to $\rho_s$.
At a minimum any further movement in the direction of the current
step increases the error metric; the gradient direction must
be perpendicular to the current step.  In other words the current step
and the next step become orthogonal:
\begin{eqnarray}
\nonumber
\tfrac{\partial}{\partial \delta}\eps^2_m(\rho+\delta \Delta \rho)&=&
\left \langle
\Delta \rho |\Ps [\bm I-\Pm] \left (\rho+\delta \Delta \rho_s \right )
\right \rangle_r\,,\\
0&=&\left \langle
\Delta \rho_s | [\bm I-\Pm] \left (\rho+\delta \Delta \rho_s \right )
\right \rangle_r\,,
\label{eq:crossER_real}
\end{eqnarray}
where $\langle \bm x|\bm y \rangle_r=\Re \left (
\bm x^{\dagger}\cdot \bm y \right )$.  The line search
algorithm can use $\epsilon^2_m$, and/or its derivative in
Eq. (\ref{eq:crossER_real}). This optimization should be
performed in reciprocal space, where the modulus projector is a
diagonal operator and is fast to compute (Eq. (\ref{eq:modulus})), while the
support projection requires two Fourier transforms:
\begin{eqnarray} 
\tPs &=&{\cal F} \bm P_s {\cal{F}}^{-1}.
\end{eqnarray}

The steepest descent method is known to be inefficient in the presence
of long narrow valleys, where imposing that successive steps be
perpendicular causes the algorithm to zig-zag down the valley.  This
problem is solved by the non-linear conjugate gradient method
\cite{hestenes, numrec, powell, polak1, polak2, fletcher, fletcher2}.  
Instead of moving in the direction of steepest descent $\Delta
\rho_s$, we move in the {\it conjugate} direction $\Lambda
\rho_s$:
\begin{equation}
\Lambda \rho_s^{(n)}=
\begin{cases}
\Delta \rho_s^{(n)} &\text{ if $n=1$}\,,\\
\Delta \rho_s^{(n)}+\gamma_s
\Lambda \rho_s^{(n-1)} &\text{otherwise,}
\end{cases}
\label{eq:cg}
\end{equation}
with $\gamma_s$ given by the Polak--Ribi\`ere method \cite{polak1}:
\begin{equation}
\label{eq:conjgamma}
\gamma_s= \tfrac{\left \langle \Delta \rho_s^{(n)}|\Delta \rho_s^{(n)}-\Delta \rho_s^{(n-1)} \right\rangle_r}{\|\Delta \rho_s^{(n-1)} \|^2} \,,
\end{equation}
and forced to be positive: $\gamma=\max(\gamma_s,0)$  to improve its reliability.
The presence of local minima shown in the previous chapters, however,
will cause stagnation of steepest and conjugate gradient methods,
preventing global convergence (Fig. \ref{G:CG}).

\subsection*{Feedback  and the saddle-point problem}

The ability to escape local minima demonstrated by input-output 
feedback-based algorithms (Fig. \ref{G:HIO}) makes them superior to the
methods based on simple gradient minimization of the error. 
However, as in the ER algorithm, the
step length is not optimized, the algorithm keeps moving in the same
direction for several steps, and sometimes overshoots. Combining the
ideas of the conjugate gradient or the steepest descent methods and IO
feedback could considerably speed-up convergence. Given the lagrangian
${\cal L}$ defined as the difference between the two errors:
\begin{eqnarray}
{\cal L}(\rho)&=&\eps^2_m(\rho)-\eps^2_s(\rho)\,,
\end{eqnarray}
using equations (\ref{eq:gradient}) and (\ref{eq:gradeps_s}) we obtain the gradient:
\begin{eqnarray}
\nabla {\cal L}(\rho)&=&2[\Ps-\Pm] \rho\,.
\end{eqnarray}
The step $\Delta \rho$ used in HIO (Eq. \ref{eq:HIO})  
can be expressed in terms of this gradient $\nabla{\cal L}$:
\begin{eqnarray}
\nonumber
\Delta \rho&=&\rho^{(n+1)}-\rho^{(n)}\,\\
\nonumber
&=&\{\Ps [\Pm -\bm I]-\beta \Pns \Pm\}\rho\,,\\
\nonumber
&=&\{\Ps [\Pm -\Ps]-\beta \Pns [\Pm-\Ps]\}\rho\,,\\
\label{eq:hio_grad_step}
	       &=& \{ -\Ps +\beta \Pns \} \half \nabla {\cal L}(\rho)\,.
\end{eqnarray}
HIO/HPR/ASR algorithms move toward 
the minimum of ${\cal L}$ in the subspace $\rho_s$, and the maximum in
the subspace $\rho_\us$, using a reduced gradient
optimization strategy, where the step is proportional to the gradient
but with one sign reversal (Eq. (\ref{eq:hio_grad_step})). In
other words, they seek the saddle point:
\begin{equation}
\min_{\rho_s} \max_{\rho_\us} {\cal L}(\rho_s+\rho_\us)\,.
\end{equation}

 Min-max or saddle-point
problems arise in fields as various as game theory, economics,
physics, engineering, and primal--dual optimization methods.  Function
minimization is easier than saddle-point optimization because a simple
function evaluation can tell us if a new point is better than the
previous one.  The saddle can be higher or lower than the current value,
although the direction toward the saddle is indicated by the two
gradient components.  One option is to alternate minimization in
the direction $\rho_s$ and maximization in the direction $\rho_\us$ of
${\cal L}$. Such a strategy is similar to alternating HIO and ER
algorithms and can be performed using off-the-shelf optimization
routines, but it can be slow. 
Optimization of the step length, a multiplicative factor $\delta$, 
 is obtained by increasing $\alpha$ until the current and next search directions
become perpendicular to one another (Fig. \ref{G:HIOSD}):
\begin{eqnarray}
\nonumber
\left \langle
\Delta \rho|
[ \Ps -\beta \Pns]\nabla {\cal L}
(\rho+\delta \Delta \rho)
\right \rangle_r&=&0\,,\\
\left \langle
\Delta \rho|
\{ \Ps[\Pm-\bm I ] -\beta \Pns \Pm\}
(\rho+\delta \Delta \rho)
\right \rangle_r&=&0\,.
\label{eq:saddle1d}
\end{eqnarray}
In analogy to the conjugate gradient method, one could substitute
the search direction $\Delta \rho$ with $\Lambda \rho$, as in
Eq. (\ref{eq:cg})  (Fig. \ref{G:HIOCG}). 
 A more robust strategy involves replacing
the one-dimensional search with a two-dimensional optimization of the
saddle point (Fig. \ref{G:HIOSD2}):
\begin{eqnarray}
\nonumber
\min_{\alpha}\max_{\beta} \psi(\alpha,\beta)\,,\\
\psi(\alpha,\beta)=
{\cal L}
(\rho+\alpha \Delta\rho_s+\beta \Delta\rho_\us)\,.
\label{eq:minmaxab}
\end{eqnarray}
  Once the 2D min--max problem is solved, the new directions can be obtained
by following the conjugate gradient scheme (Fig. \ref{G:HIOCG2s}).

\section{Conclusions}
\begin{figure}
\fbox{	\includegraphics[width=0.2\textwidth,clip]{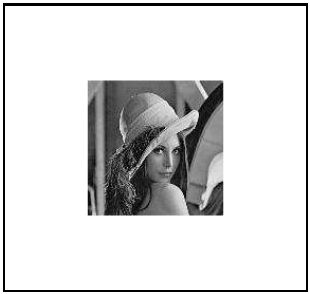}	}
\caption{\label{fig:lena}
Test figure used for benchmarking. The object of $128^2$ elements is
surrounded by empty space. The whole image has $256^2$ elements.
The Fourier transform of this image provides the data set, and its
area defines the support}
\end{figure}

\begin{figure}
	\includegraphics[width=0.4\textwidth]{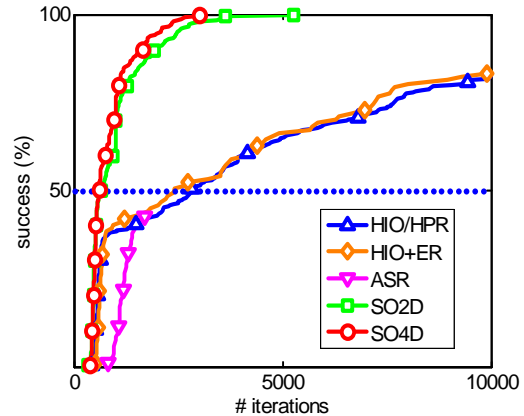}	
	\caption{
Percentage of successful reconstructions over many tests starting from random phases 
as a function of number of iterations. The support is the only constraint. Positivity and reality
are not enforced, and the support is {\it loose}: it is larger than the object by one additional
row and column  \label{fig:benchmark}}
\end{figure}

Lensless imaging owes its success as an effective tool to observe
nanoscale systems to the advances made in phase-retrieval algorithms.
The new instruments replacing lenses are the iterative projection
algorithms for phase retrieval.  These algorithms can be grouped in
two categories: (1) local minimizers such as ER, SF,
steepest descent and conjugate-gradient methods, with Solvent Flip
having some moderate ability to escape local minima
\cite{abrahams:1996}. (2)  more global minimizers
such as HIO, DM, ASR, HPR which use a feedback to reach the solution.
RAAR and ER+HIO fall somewhere in between the two categories,
depending on an adjustable parameter.  A simple benchmark is shown for
comparison in Fig. \ref{fig:lena} and summarized in Table
\ref{tab:benchmark}. The test consisted in solving a phase-retrieval
problem without assuming positivity (nor reality) of the object, and
the support region was slightly larger than the object, and was
repeated 100 times for each algorithm.  Many algorithms surprisingly
failed, and only the ones shown in Fig. \ref{fig:benchmark}
succeeded. HIO appears to be the most effective algorithm, and it is
significantly improved in terms of speed and reliability when the
two-dimensional step size optimization (SO2D), as described in
Eq. (\ref{eq:minmaxab}) is applied. Further improvements in
reliability are achieved by performing a saddle-point optimization in
a 4-dimensional space of two successive steps (SO4D).  Minimization
algorithms, although not very powerful at solving the phase problem,
can be used to polish-up a solution, improving the values of the error
metric considerably.  The algorithms described here use as prior
knowledge the support region. Algorithms that use a simple threshold
to replace the support \cite{Millane:1996, chargeflip} or more
sophisticated support refinement \cite{Marchesini:2003} have not been
discussed. Various projection algorithms combined with some form of
threshold have produced remarkable reconstructions of single isolated
objects (HIO and RAAR \cite{Chapman:2006}, DM \cite{Shapiro:2005}), as
well as single and powder crystals (SF and HIO with support refinement \cite{chargeflip,
spencecrystal, spencepowder}), but a full comparison of the algorithms
behavior applied to this type of constraint have not been discussed.

\begin{table}
\caption{
Benchmark of various algorithms
 \label{tab:benchmark}
}
\begin{tabular}{|l|l|l|}
\hline
Algorithm & No. of iterations  for& success after \\
          &  50\% success &  10000 iterations\\
\hline
HIO/HPR& 2790 & 82\% \\
HIO/HPR+ER& 2379& 82.6\% \\
ASR& 1697\footnote{42\% success, the algorithm either reconstruct
in a limited number of iterations or never}& 42\% \\
SO2D& 656 & 100\%\\
SO4D& 605 & 100\%\\
Others & $> 10000$ & 0\%\\
\hline
\end{tabular}
\end{table}

\begin{acknowledgments}
This work was performed under the auspices of the U.S. Department of 
Energy by the Lawrence Livermore National Laboratory under Contract 
No. W-7405-ENG-48 and the Director, Office of Energy Research. 
This work was partially funded by the  National Science Foundation through the Center
 for Biophotonics. 
The Center for Biophotonics, a National Science Foundation
 Science and Technology Center, is managed by the University of 
California, Davis, under Cooperative 
Agreement No. PHY0120999. The author acknowledge useful discussions with H. N. Chapman, M. R. Howells, J.~C.~H. Spence and D.~R.~Luke.
\end{acknowledgments}

\printfigures


\end{document}